%% file: multipol.tex
\documentclass[preprint,12pt,3p]{elsarticle}

\usepackage{tikz}
\usepackage{matlab-prettifier}
\usepackage{graphicx}
\usepackage{makecell}
\usepackage{xfrac}
\usepackage{textcomp}
\usepackage{parskip}
\usepackage{adjustbox}
\usepackage{setspace}
\usepackage{lscape}
\usepackage[backref=page]{hyperref}

\usepackage{booktabs}
\usepackage{graphicx}  % for \resizebox
\usepackage{threeparttable} 
\usepackage{siunitx}        
\usepackage{array}          
\usepackage{makecell}      
\usepackage{multirow}
\usepackage{float}
\renewcommand{\arraystretch}{1.1}    
\newcommand{\sym}[1]{\rlap{#1}}
\newcommand{\textoverline}[1]{$\overline{\mbox{#1}}$}
\hypersetup{ 
backref=true, % link backreferences (i.e. citations to their list)
bookmarks=true, % show bookmarks bar?
pdftoolbar=true, % show Acrobats toolbar?
pdfmenubar=true, % show Acrobats menu?
pdffitwindow=false, % window fit to page when opened
pdfstartview={FitH}, % fits the width of the page to the window
pdftitle={}, % title
pdfauthor={}, % author
pdfsubject={}, % subject of the document
pdfkeywords={}{}, % list of keywords
pdfnewwindow=true, % links in new window
colorlinks=true, % false: boxed links; true: colored links
linkcolor=BlueViolet, % colour of internal links
citecolor=maroon, % colour of links to bibliography
urlcolor=blue,
}
\usepackage{comment}
\usepackage{subfig}
\usepackage{caption}
\usepackage{longtable}
\usepackage{amssymb}
\usepackage{amsmath}
\usepackage{gensymb}
\usepackage{siunitx}

\usepackage{natbib}
\biboptions{comma,round}
\setcitestyle{authoryear}

\begin{document}
%\begin{spacing}{1.5}
\begin{frontmatter}

\title{Interactions Between Multiple Environmental Markets: Addressing Contamination Bias in Overlapping policies}

\author[label2]{Tiantian Yang}
\author[label1,label3,label4,label5,label6]{Richard S.J. Tol\corref{cor1}%\fnref{label7}
}

\address[label2]{Jinhe Center for Economic Research, Xi'an Jiaotong University, Xi'an, China}
\address[label1]{Department of Economics, University of Sussex, Falmer, United Kingdom}
\address[label3]{Department of Spatial Economics, Vrije Universiteit, Amsterdam, The Netherlands}
\address[label4]{Tinbergen Institute, Amsterdam, The Netherlands}
\address[label5]{CESifo, Munich, Germany}
\address[label6]{Payne Institute for Public Policy, Colorado School of Mines, Golden, CO, USA}

\cortext[cor1]{Jubilee Building, BN1 9SL, UK}
%\fntext[label7]{}

\ead{r.tol@sussex.ac.uk}
\ead[url]{http://www.ae-info.org/ae/Member/Tol\_Richard}

\begin{abstract}
To address the dual environmental challenges of pollution and climate change, China has established multiple environmental markets, including pollution emissions trading, carbon emissions trading, energy-use rights trading, and green electricity trading. Previous empirical studies suffer from known biases arising from time-varying treatment and multiple treatments. To address these limitations, this study adopts a dynamic control group design and combines Difference-in-Difference (DiD) and Artificial Counterfactual (ArCo) empirical strategies. Using panel data on A-share listed companies from 2000 to 2024, this study investigates the marginal effects and interactive impacts of multiple environmental markets implemented in staggered and overlapping phases. Existing pollution emissions trading mitigates the negative effects of carbon emission trading. Carbon trading suppresses (improves) financial performance (if implemented alongside energy-use rights trading). The addition of energy-use rights or green electricity trading in regions already covered by carbon or pollution markets has no significant effects.\\
\textit{Keywords}: Multiple environmental markets; Policy interactions; Marginal abatement cost; Contamination bias; Artificial Counterfactual; Difference-in-Difference\\
\medskip\textit{JEL codes}: Q54
\end{abstract}

\end{frontmatter}

\section{Introduction}
The emissions of greenhouse gases and air pollutants are two of the most critical environmental challenges facing the world today, exerting severe impacts on public health, society, economy, and labor \citep{tol1994damage, chay2005does, tol2018economic, herrnstadt2021air, chen2022effect}. In response, many countries have introduced multiple environmental policies that continue to evolve. This complicates policy evaluation. We address this problem for China, studying four permit trading schemes that were rolled out over time in some but not all provinces.

A substantial body of literature has pointed to potential interactions between environmental policies, noting that such interactions may be synergistic, neutral, or conflicting \citep{rogge2016policy, wilts2019policy, van2021designing}. Especially for the carbon emission trading system, numerous studies show the interaction between EU-ETS and Kyoto Protocol flexibility mechanisms \citep{hintermann2019linking}, electricity‐market structures \citep{bersani2022ets}, renewable‐energy certificates \citep{wu2024interplay, morthorst2001interactions}, and other renewable‐energy incentives \citep{proencca2020synergies, del2007interaction, fischer2010combining}. In China, recent research has examined the interaction between carbon emission trading and energy-use rights trading \citep{li2019synergy, sun2024compliance}, pollution emission trading\citep{sun2023synergy, zhu2023coeffect}, and green electricity trading \citep{wei2023interaction, wang2021coordination, zhang2023assessing}. If such interactions are not properly accounted for in policy evaluations, the estimated effects may be biased or misinterpreted.

Nonetheless, most empirical studies rely heavily on the Difference-in-Differences (DiD) approach to estimate average treatment effects of individual policies \citep{chen2022environmental, luan2025impact, wang2024impact, tang2023does}. This approach faces two key limitations. First, staggered policy implementation biases two-way fixed effect (TWFE) DiD estimators \citep{callaway2021difference,goodman2021difference,borusyak2024revisiting}. Second, even when policies are independent, non-linear dependencies between covariates can result in contamination bias if there are multiple treatments, undermining causal inference \citep{goldsmith2024contamination}. Therefore, current methods often fall short in accurately isolating the effect of one policy when others are simultaneously in place.

To address the identification challenges posed by time-varying treatment \citep{callaway2021difference} and multiple treatments \citep{goldsmith2024contamination}, we apply phase-specific and region-specific DiD estimations by excluding the contaminated control groups. We further introduce a more general method\textemdash Artificial Counterfactual (ArCo, \cite{carvalho2018arco})\textemdash to supplement and validate the DiD results. The DiD relies on the parallel trends assumption, whereas ArCo uses the treated units’ pre-treatment trajectory to predict their counterfactual outcomes, allowing for inference even under non-parallel trends. We thus avoid the biases due to time-varying treatment and multiple treatments, as well as the biases from interactions between the treatments. The joint application of DiD and ArCo enhances robustness of causal analysis in complex policy environments.

China ranked 156th among 180 countries in the 2022 Environmental Performance Index (EPI) \citep{yale_epi2024}. China accounts for approximately $1/3$ of global carbon dioxide emissions \citep{iea2022china} and hosts the world’s largest carbon trading market by coverage. More importantly, China is now entering a critical period of transition from fragmented policymaking to integrated governance. The 2022 national policy on building a unified market explicitly calls for the consolidation of environmental markets, and the 2023 National Conference on Ecological and Environmental Protection emphasizes the importance of policy coordination and multi-pollutant governance. Against this backdrop, this study aims to systematically analyze the effects of environmental markets in China, identify the synergies and frictions within the ongoing institutional integration, and provide valuable policy implications for other high-emission economies. 

There is an ongoing debate regarding the effects of environmental markets on companies’ financial performance. The key controversy is whether these market mechanisms unduly increase companies’ financial burdens\citep{lanoie1998firms} or, if well designed, spur innovation to deliver the dual benefits of environmental protection and economic performance \citep{porter1996america}. 

Prior empirical studies suggest that energy-use rights \citep{wang2024impact, wang2025towards} and green electricity trading \citep{tang2023does} enhance companies’ financial performance, but there are mixed results for pollution emission trading \citep{chen2022environmental, liu2022so2} and carbon emission trading \citep{luan2025impact, li2025impact}. Replicating these studies’ empirical strategies, we find similar results. However, once contaminated control groups are excluded, significant effects only appear for companies in non-pilot regions. For companies previously subject to pollution or carbon trading, additional policies show no significant effect on companies' financial performance. This highlights the limited marginal benefit of overlapping policies and underscores the need for integrated environmental market design.

Carbon emissions trading reduces companies’ financial performance, whereas existing pollution emissions trading mitigates this negative effect. Specifically, the estimated effect changes from –0.618\% to a statistically insignificant negative value under the DiD approach, and from –1.044\% to a statistically insignificant positive value under the ArCo approach. Moreover, the simultaneous implementation of energy-use rights trading further offsets the adverse impact of carbon markets on firm performance. The estimated effect increases from –0.618\% to 1.172\% under DiD, and from –1.044\% to 1.750\% under ArCo, suggesting that overlapping environmental markets may offer opportunities for cross-market arbitrage. 

This paper contributes to the literature in three ways: (1) unlike prior studies that focus on one or two environmental markets, this paper systematically examines four major environmental markets\textemdash pollution trading, carbon trading, energy-use rights trading, and green electricity trading\textemdash enabling a unified analysis of their interactions and combined effects on companies’ abatement costs; (2) this paper develops a dynamic identification strategy using phase-specific and region-specific DiD estimations to address known biases from time-varying treatment and multiple treatments and unknown biases due to their interaction, thereby enabling estimation of marginal policy effects and more accurate estimation of policy effects; (3) this paper integrates DiD and ArCo methodologies to enhance the robustness of causal inference. While DiD relies on the parallel trends assumption, ArCo constructs counterfactual from within-group trajectories and avoids contamination bias, complementing the DiD framework under complex policy environments.

The paper proceeds as follows. Section \ref{sc:data} presents the data in the context of energy and environmental regulation in China. Section \ref{sc:strategy} introduces the empirical strategy including DiD and ArCo. Section \ref{sc:results} presents the empirical results, including the the main results of DiD and ArCo, the comparison between the two methods and reason analysis. Section \ref{sc:discussion} replicate the results of the existing literature and compare the results based on different empirical strategy. Section \ref{sc:conclusions} concludes.    

\section{Context and data}
\label{sc:data}
\subsection{The environmental markets in China}
\begin{figure}[H]
    \centering
    \includegraphics[width=1.0\linewidth]{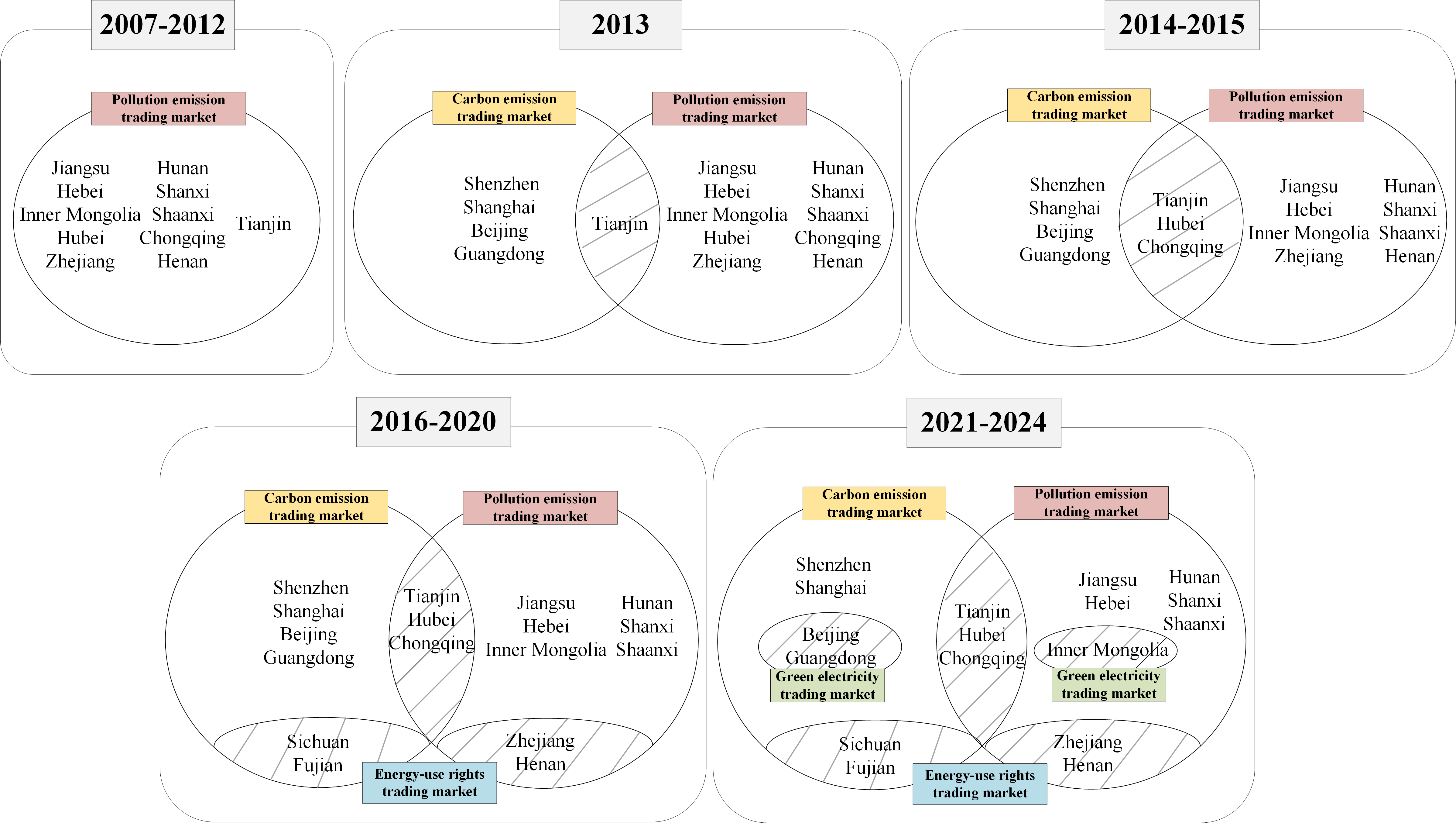}
    \caption{The implementation timeline of the four environmental markets.}
    \label{fig:implement process}
\end{figure}

There are four main environmental markets in China: air pollution emission trading, carbon emission trading, energy-use rights trading, and green electricity trading. These markets have been implemented through phased pilot programs at different times and across different, sometimes overlapping, provinces. Figure~\ref{fig:implement process} illustrates the timeline and geographic distribution of these four environmental market pilots.

As illustrated in Figure~\ref{fig:implement process}, pollution emission trading was implemented across eleven pilot regions between 2007 and 2024. Carbon emission trading was first launched in 2013 in five pilot regions, with Tianjin geographically overlapping with an existing pollution emission trading pilot. Between 2014 and 2015, two additional carbon trading pilots\textemdash Hubei and Chongqing\textemdash were introduced, both fully overlapping with regions already covered by pollution emission trading.

From 2016 to 2020, energy-use rights trading was piloted in four regions, and carbon trading expanded to two additional provinces. Sichuan and Fujian served as simultaneous pilots for both carbon and energy-use rights trading, while Zhejiang and Henan overlapped entirely with existing pollution emission trading pilots.

During 2021–2024, green electricity trading was initiated in Beijing, Guangdong, and Inner Mongolia. Among them, Beijing and Guangdong overlapped with carbon emission trading pilots (but not pollution and energy trading), while Inner Mongolia overlapped with pollution emission trading but not carbon and energy-use rights trading.

\subsection{The contamination bias in multiple environmental policies background}
The pollution emissions trading launched in 2007, the energy-use rights trading in 2016, and the green electricity trading in 2021 are analyzed using single-period DiD designs, while the carbon emissions trading introduced in 2013, 2014, and 2016 is analyzed using staggered DiD. In staggered adoption settings, TWFE DiD regressions suffer from negative weighting problems \citep{callaway2021difference, goodman2021difference, borusyak2024revisiting}. Moreover, \cite{goldsmith2024contamination} point to another problem: contamination bias occurs when additive adjustments to covariates fail to capture non-linear relationships between a given treatment and other treatments or covariates. As a result, linear regression may incorrectly assign a non-zero fitted probability to a given treatment when another treatment has already been implemented. Prior research has largely emphasized a single policy's effects \citep{chen2022environmental, luan2025impact,wang2025towards}, although there are other policies implemented at the same time. 

Therefore, we construct a phased, multi-level control design based on the temporal differences and spatial overlap of multiple environmental pilots. This design prevents the conflation of impacts from different markets and enables the identification of the marginal effects of each pilot, where appropriate conditional on another pilot. 

Consider Tianjin in 2013 as an example, There are two treatments: tradable permits for air pollution and carbon dioxide. As both originate largely from the combustion of fossil fuels, these policies interact with one another. We could estimate a two-way fixed-effect difference-in-differences model but this would suffer from the known biases due to time-varying treatment \citep{callaway2021difference} and multiple treatments \citep{goldsmith2024contamination} and the unknown bias due to their interaction.

\include{Tables.tex/identification}

Therefore, we instead restrict the control group to those provinces that have pollution trade but no carbon trade (Jiangsu, Hebei, Inner Mongolia, Hubei, Zhejiang, Hunan, Shanxi, Shaanxi, Chongqing, Henan) and years after the implementation of pollution emission trading in 2007 (Panel (3) in Table~\ref{tab:identification}). This identifies the impact of the carbon market \emph{on top of} the pollution market.

In order to identify the impact of the market in CO\textsubscript{2} emission permits proper, we compare companies in Shenzhen, Shanghai, Beijing, Guangdong (but not in Tianjin) to companies in the other provinces that have neither carbon nor air pollution markets (Panel (2) in Table~\ref{tab:identification}).

Panel (4) in Table~\ref{tab:identification} illustrates a different setting. Tianjin was designated as a carbon trading pilot in 2013, followed by Hubei and Chongqing in 2014. Accordingly, a staggered DiD is required, and careful attention must be paid to the issue of negative weighting, as discussed above.

In 2016, Sichuan and Fujian simultaneously became pilots for both carbon emission trading and energy-use rights trading. As companies in these provinces had not previously been subject to carbon trading, it is not feasible to compare companies in Sichuan and Fujian with those in Shenzhen, Shanghai, Beijing, and Guangdong, which had already participated in carbon markets. Consequently, we cannot estimate the additional effect of energy-use rights trading relative to existing carbon trading in these two provinces. Instead, companies in Sichuan and Fujian can only be compared to those in provinces without any environmental markets (as shown in Panel (5) of Table~\ref{tab:identification}). However, this comparison does not allow us to distinguish whether the observed effects are driven by carbon trading, energy-use rights trading, or a combination of both, relative to non-pilot regions.

In contrast to companies in Sichuan and Fujian, companies in Zhejiang and Henan were additionally subject to energy-use rights trading from 2016. The additional effect of energy-use rights trading on top of existing pollution trading pilots can be estimated (as shown in Panel (6) of Table~\ref{tab:identification}).

In Panel (7), Shenzhen, Shanghai, Beijing and Guangdong have all implemented the same emission trading scheme since 2013, with Beijing and Guangdong additionally implementing green electricity trading in 2021. So the additional electricity effect on top of carbon emission trading can be identified. Similarly, in Panel (8), the incremental effect of green electricity trading relative to existing pollution emission trading can be estimated.

\subsection{Data selection and description}
\subsubsection{Rationale for selecting Return on Assets (ROA)}

This study adopts ROA as the primary outcome variable to assess the economic impact of environmental markets. ROA is a standard measure of profitability\textemdash profits relative to assets. ROA captures environmental compliance costs. ROA is widely used in the existing literature, ensuring comparability with prior studies.

Alternative financial indicators, such as Return on Equity (ROE), Price-to-Earnings ratio (P/E), Price-to-Book ratio (P/B), and Tobin’s Q, are less suited \citep{luan2025impact}. ROE is sensitive to capital structure and can be inflated through leverage, while Tobin’s Q is driven by investor sentiment and macroeconomic fluctuations, making it unsuitable for evaluating short-term regulatory impacts. Similarly, valuation-based ratios like P/E and P/B are shaped more by market expectations than by companies’ actual cost structures.

In contrast, ROA is less affected by market sentiment and financial leverage, offering a clearer attribution of regulatory impacts on real economic activity. Under environmental markets, companies often face rising capital expenditure (e.g., investment in clean technologies) and profit compression due to carbon pricing or compliance penalties—both of which are included in ROA. Thus, ROA provides a theoretically sound and empirically consistent measure of regulatory cost exposure across companies with heterogeneous financing and market conditions.

\subsubsection{Control Variable Selection}
To mitigate potential confounding effects and more accurately identify the impact of environmental markets on firm performance, this study incorporates a set of control variables grounded in the empirical literature on corporate finance and industrial organization \citep{luan2025impact, huang2025has, chen2024emission, dechezlepretre2023joint}. These variables capture company-specific characteristics and industry structure that may independently influence profitability and regulatory responsiveness.

The Herfindahl-Hirschman Index (HHI) proxies industry concentration, with higher values indicating reduced competition and greater capacity for dominant companies to pass on compliance costs. Company Age (AGE), measured as the logarithm of years since establishment, reflects organizational maturity and adaptive capacity. Employment Size (EMP), measured as the logarithm of total employees, captures company scale and complexity, which may affect both adjustment costs and compliance capacity. The Operational Capital to Current Assets Ratio (OCCAR) reflects companies’ investment strategies and liquidity management, with higher values indicating greater commitment to long-term assets and potential resilience to compliance costs. The Debt-to-Equity Ratio (DER) captures financial leverage, influencing companies’ risk exposure and strategic responses to regulation.

Variable definitions and calculation methods are detailed in Table~\ref{tab:variable_description}.

\input{Tables.tex/Table-data_description}

These controls account for key dimensions of company heterogeneity\textemdash market position, maturity, scale, capital allocation, and financial structure\textemdash ensuring more credible identification of regulatory effects.

\subsubsection{Data source and descriptive statistics}

Existing research on environmental markets typically adopts one of two strategies. The first operates at the regional level, treating prefecture-level cities or provinces that implemented pilot programs as the treatment group \citep{zhou2022assessing}. The second focuses on company-level analysis, using all A-share listed companies as the sample and classifying those located in pilot regions as the treatment group \citep{liu2022so2,wang2024impact}. Some studies further refine this approach by focusing on specific industries and selecting companies within pilot regions and within the target industry as the treatment group \citep{chen2022environmental,tang2023does}.

However, due to limited disclosure regarding company-level participation in pollution emission trading and energy-use rights trading, no existing studies have been able to identify the actual participants in these markets. In the case of carbon emission trading, some scholars have used the subset of A-share listed companies included in official lists of key emission-control companies as proxies for participation \citep{luan2025impact}. Yet this method presents significant limitations. As \citet{huang2025has} observe, “the pilot firm list consists of more than 2000 entities, among which only 78 are A-share listed companies.” Moreover, inclusion in these lists does not guarantee actual participation in trading activities, nor does it rule out the possibility that other companies were affected by the trading scheme. These issues result in a substantially reduced sample size and introduce potential selection bias.

To address these challenges and ensure data availability, empirical consistency, and identification credibility, this study adopts a widely used empirical strategy. We first use the full sample of A-share listed companies, treating those located in pilot regions as the treatment group. As a robustness check, we further use the subsample of regulated industries based on the local governments' official documents and repeat the analysis with that subsample to validate the results in an industry-specific context.

The data is sourced from the China Stock Market and Accounting Research Database (CSMAR). The companies marked with ST or ST* are excluded. The dependent and independent variables are truncated at the 1\% and 99\% quantiles. Our data set includes all listed A-share companies in China and spans from 2000 to 2024. The descriptive statistics are shown in Table~\ref{Table_data_statistics}. 

\input{Tables.tex/Table_data_statistics}

Table~\ref{tab:data_group_statistics} presents the mean values and standard deviations of ROA of each panel. 

\input{Tables.tex/Table-data_group_statistics}

Table~\ref{tab:data_group_statistics} reports the summary statistics of ROA across different treatment and control panels. Companies in pollution emission trading regions exhibit higher ROA than those in non-pilot regions, while companies in carbon emission trading regions show slightly lower ROA. The addition of energy-use rights trading is associated with higher ROA, suggesting potential complementarities with existing markets. In contrast, the effect of adding green electricity trading on ROA appears heterogeneous

These preliminary observations may be influenced by confounding factors such as enterprise industry classification, operational scale, and other covariates. Subsequent analyses will systematically control for these variables to rigorously investigate the dynamic policy effects and cumulative interactions of pilot implementations on corporate ROA.

\section{Empirical strategy}
\label{sc:strategy}
\subsection{Model specification}
The three approaches in policy evaluation\textemdash Synthetic Control (SC), Difference-in-Differences (DiD), and Artificial Counterfactual (ArCo)\textemdash differ in their assumptions, counterfactual construction, and ability to capture dynamic policy effects \citep{carvalho2018arco}.

While SC is theoretically appealing, it is not suitable for the present study. SC constructs a synthetic control group as a weighted average of untreated units, using non-negative weights that sum to one ($\hat \Delta_{SC} = \frac{1}{T-T_0+1} \sum_{t=T_0}^T (y_{1t} - \hat y_{1t}),\hat{y}_{1t} = \sum_{i=2}^{n} w_i\, y_{it}, w^* = \arg\min_{w \ge 0, \;\sum w = 1} \left\| \bar{z}_1 - w^\top \bar{z}_0 \right\|_V$). However, it relies on pre-intervention averages and discards time-series dynamics, which limits its ability to capture staggered and cumulative policy effects. Moreover, SC is only applicable to balanced panel data, whereas our sample includes companies that entered or exited the market mid-period due to Initial Public Offerings (IPOs), delistings, or bankruptcies. As a result, the panel is unbalanced, making SC unsuitable for this analysis.

DiD compares average outcomes between treated and control groups before and after policy implementation ($\hat \Delta_{DID} = \left[(\bar Y_{post}^{treat} - \bar Y_{pre}^{treat}) - (\bar Y_{post}^{control} - \bar Y_{pre}^{control}) \right]$), assuming parallel trends in the absence of treatment. While DiD provides intuitive and widely accepted estimates of average treatment effects, its reliability hinges on having sufficiently long post-treatment windows and no interference from overlapping policies. Given the complex and phased introduction of emissions trading schemes\textemdash including pollution emission trading (2007), carbon emission trading (2013–2014), energy-use rights trading (2016), and green electricity trading (2021)\textemdash DiD’s assumptions may be difficult to satisfy, particularly in later policy phases.

By contrast, ArCo offers a more general and flexible framework. It does not rely on the parallel trends assumption and allows for nonparametric functional forms. It can construct counterfactuals even when treated and control units exhibit divergent pre-treatment trends, by employing nonparametric weighting ($ \hat \Delta_T = \frac{1}{T-T_0+1} \sum_{t=T_0}^T \hat \delta_t, \hat \delta_t = y_t -M(Z_{0t}, \hat \theta_{T_1}), M(Z_{0t},\hat\theta)\;=\;\bigl(x_{1t}'\hat\theta_1,\dots,x_{qt}'\hat\theta_q\bigr)'$). Moreover, ArCo preserves the full temporal structure of the data, capturing dynamic responses over time and enabling formal statistical inference. This is particularly valuable in evaluating environmental markets with shorter post-intervention windows or policy overlaps (e.g., green electricity trading). 

Therefore, we focus on DiD and ArCo, which are better aligned with the structure of China’s environmental markets rollout. Both methods are applied within a recursive framework, allowing for dynamic assessment across sequentially implemented policies. 

In this study, DiD is used for estimating the average treatment effects. However, to address potential identification biases commonly associated with DiD\textemdash including the known biases due to time-varying treatment \citep{callaway2021difference} and multiple treatments \citep{goldsmith2024contamination} and the unknown bias due to their interaction— and to account for the possibility of failing the parallel trends assumption, the ArCo is adopted as supplementary analytical tool. This dual-method approach allows for cross-validation of results and mitigates the risk of biased conclusions driven by the limitations of a single specification.

\subsection{Methodology}
\subsubsection{Difference-in-Difference (DiD)}
The standard TWFE DiD model is specified as follows:
\begin{equation}
    ROA_{it} = \alpha + \beta Treat_i \times Post_t + \gamma X_{it} + \mu_i + \lambda_t + \varepsilon_{it} \nonumber
\end{equation}
Where $ROA_{it}$ denotes the return on assets of company $i$ at time $t$; 
$Treat_i$ is a dummy variable indicating whether company $i$ belongs to the treatment group; 
$Post_t$ is a post-treatment time dummy; 
$X_{it}$ represents a set of control variables; 
$\mu_i$ and $\lambda_t$ are company and year fixed effects, respectively;
$\varepsilon_{it}$ is the error term. 
The coefficient $\beta$ captures the average treatment effect.

However, the validity of DiD relies heavily on the parallel trends assumption and requires sufficiently long post-treatment periods to accurately estimate dynamic effects. Given the complexity of overlapping policy treatments and short implementation windows of newer markets (e.g., green electricity trading from 2021 onwards), DiD yields biased estimates in such contexts.

Consider Tianjin in 2013 (cf. Figure~\ref{fig:implement process}). There are two treatments: tradable permits for air pollution and carbon dioxide. As both originate largely from the combustion of fossil fuels, we cannot assume that these policies do not interact with one another. We could estimate
\begin{equation}
    ROA_{it} = \alpha + \beta^A Treat^A_i \times Post^A_t + \beta^B Treat^B_i \times Post^B_t+ \gamma X_{it} + \mu_i + \lambda_t + \varepsilon_{it} \nonumber
\end{equation}
but this would suffer from the known biases due to time-varying treatment \citep{callaway2021difference} and multiple treatments \citep{goldsmith2024contamination} and the unknown bias due to their interaction. Therefore, we instead restrict the sample to those provinces and years for which $Post^A_t = 1$. The parameters $\beta_B$ is then the causal impact of \emph{adding} CO\textsubscript{2} to air pollution permits (Panel (3) in Table~\ref{tab:identification}).

In order to identify the impact of the market in CO\textsubscript{2} permits, we compare companies in Beijing, Guangdong, Shanghai, and Shenzhen (but not in Tianjin) to companies in the 16 provinces that have neither carbon nor air pollution markets (Panel (2) in Table~\ref{tab:identification}).

Moreover, as noted earlier, Panel (4) involves staggered treatment adoption. Applying a standard TWFE DiD model in this context would lead to biased estimates due to negative weighting issues. Therefore, we adopt the methodology of \citet{callaway2021difference} to correct for these biases and enhance the robustness of our results. Their approach constructs comparison groups based on units that have not yet been treated in period $t$, thereby mitigating concerns over potential biases inherent in TWFE estimates.

\subsubsection{Artificial Counterfactual (ArCo)} 
To address these limitations, we employ a variant of the ArCo as a supplementary methodology \citep{carvalho2018arco}. ArCo constructs counterfactual outcomes using a predictor model estimated on treated companies, with untreated companies as explanatory variables. Unlike DiD, ArCo does not assume parallel trends and retains full time-series dynamics.

In the ArCo framework, the first stage requires estimating a flexible predictor model for the counterfactual outcome using a set of untreated peers’ covariates. Unlike the high-dimensional version that adopts regularization techniques such as \textit{LASSO}, we employ a linear regression model for estimation. This approach remains effective under moderate-dimensional settings where the number of predictors is smaller than the number of pre-treatment observations. The ArCo model is specified as follows:
\begin{equation}
    \hat{ROA}^{(0)}_{1t} = \alpha + \beta X_{1t} + \lambda \cdot \bar{ROA}_{-1,t} + \gamma \cdot \bar{X}_{-1,t} + \varepsilon_{it}
\end{equation}
where $ROA_{1t}$ denotes $ROA$ for the treated unit in period $t$ before the treated time $T_0$; $X_{1t}$ represents the company-level covariates of the treated unit; $\bar{ROA}_{-1,t}$ and $\bar{X}_{-1,t}$ denote the contemporaneous yearly averages of $ROA$ and covariates among the untreated control group; $\varepsilon_{it}$ is the idiosyncratic error term.

Note that ArCo was conceived for large \textit{T} and small \textit{N}. We therefore replaced the observations of the \emph{individual} companies in the control group with the \emph{average} over the control group \citep{bai2009panel}. This is a special case of the approach proposed by \citet{xu2017generalized}, who sees this as a generalization of Difference-in-Differences rather than Synthetic Control.

The parameters $\alpha$, $\beta$, $\lambda$, and $\gamma$ are estimated on pre-treatment data $t < T_0$. Once the coefficients are estimated, we predict the treated unit's counterfactual outcome in the post-treatment periods $t \ge T_0$. The treatment effect at each time point is computed as:
\begin{equation}
    \delta_t = ROA_{1t} - \hat{ROA}^{(0)}_{1t}
\end{equation}
And the average treatment effect over the post-treatment period is given by:
\begin{equation}
    \hat{\Delta}_T = \frac{1}{T - T_0 + 1} \sum_{t = T_0}^{T} \delta_t
\end{equation}
That is, ArCo tests whether treatment reduced the predictive skill of the model, as measured by an increased gap between the performance of treated and untreated companies. This is similar in spirit to Synthetic Control.

Compared to DiD, ArCo allows for non-parallel pre-trends and greater flexibility in counterfactual construction, while retaining a transparent model structure and interpretable parameters.

\section{The influence of environmental markets on companies' return on assets}
\label{sc:results}
\subsection{Results of DiD} 
\subsubsection{Main results}
Table~\ref{tab:DiD_main_results} reports the DiD estimates for ROA of the eight different panels. Since Tianjin implemented carbon emission trading in 2013, and Hubei and Chongqing followed in 2014, a staggered DiD approach is required in Panel (4). Given that the standard TWFE staggered DiD suffers from negative weighting, two DiD models are employed for Panel (4): the conventional TWFE DiD estimates reported in Table~\ref{tab:DiD_main_results}, and the estimates corrected following \cite{callaway2021difference} (Stata's \textsc{csdid}), presented in Table~\ref{tab:CSDID}.

\input{Tables.tex/Table-DiD_main_results}

\input{Tables.tex/Table_CSDID}

A comparison between Panels (2), (3), and (4) reveals that, relative to companies in non-pilot regions, the 2013 carbon emission trading pilots significantly reduce companies’ ROA by 0.865\%. However, this negative effect becomes statistically insignificant when compared to companies in regions already subject to pollution emission trading, as indicated by the insignificant coefficients of -0.385\% in Panel (3) and -0.121\% in Panel (4) (estimated by \textsc{csdid}). Assuming that the effect of carbon emission trading is homogeneous across pilot regions, this finding suggests that prior implementation of pollution emission trading mitigates the adverse impact of carbon trading on companies' ROA. Consequently, the marginal effect of carbon emission trading relative to existing pollution emission trading—that is, the \emph{adding} carbon effect—is statistically insignificant.

Similarly, a comparison between Panels (2) and (5) shows that the 2013 carbon emission trading pilots suppressed companies’ ROA relative to non-pilot regions. However, the simultaneous implementation of carbon emission trading and energy-use rights trading in Hubei and Sichuan in 2016 appears to have significantly enhanced companies’ ROA by 1.172\% compared to those in non-pilot regions. Nevertheless, it remains unclear whether this improvement stems from the carbon emission trading, the energy-use rights trading, or the interaction between the two. Assuming the effect of carbon emission trading is consistent across all provinces, it can be inferred that energy-use rights trading have offset or mitigated the negative impact of carbon emission trading on companies’ ROA.

Furthermore, Panels (1), (2) and (5) show that the observed policy effects\textemdash whether positive or negative\textemdash are identified only relative to companies in non-pilot regions. Combining this observation with the results from Panels (3), (4), (6), (7), and (8), it becomes evident that for regions where either carbon emission trading or pollution emission trading had already been implemented, further introduction of energy-use rights trading or green electricity trading does not significantly produce any additional impact on companies’ ROA. In other words, regardless of whether carbon emission trading (since 2013) or pollution emission trading (since 2007) had positive or negative effects on companies, newly introduced environmental markets do not further amplify or mitigate these effects.

Unlike Panels (1), (2) and (5), there are no clean subsamples to estimate the effects of energy-use rights trading or green electricity trading relative to non-pilot regions. As a result, we are unable to identify their interaction effects with carbon emission trading or pollution emission trading. The analysis can only capture the \emph{additional} energy or \emph{additional} electricity effects conditional on the existing carbon or pollution trading schemes.

\subsubsection{Event study}
\begin{figure}[H]
    \centering
    \includegraphics[width=\textwidth]{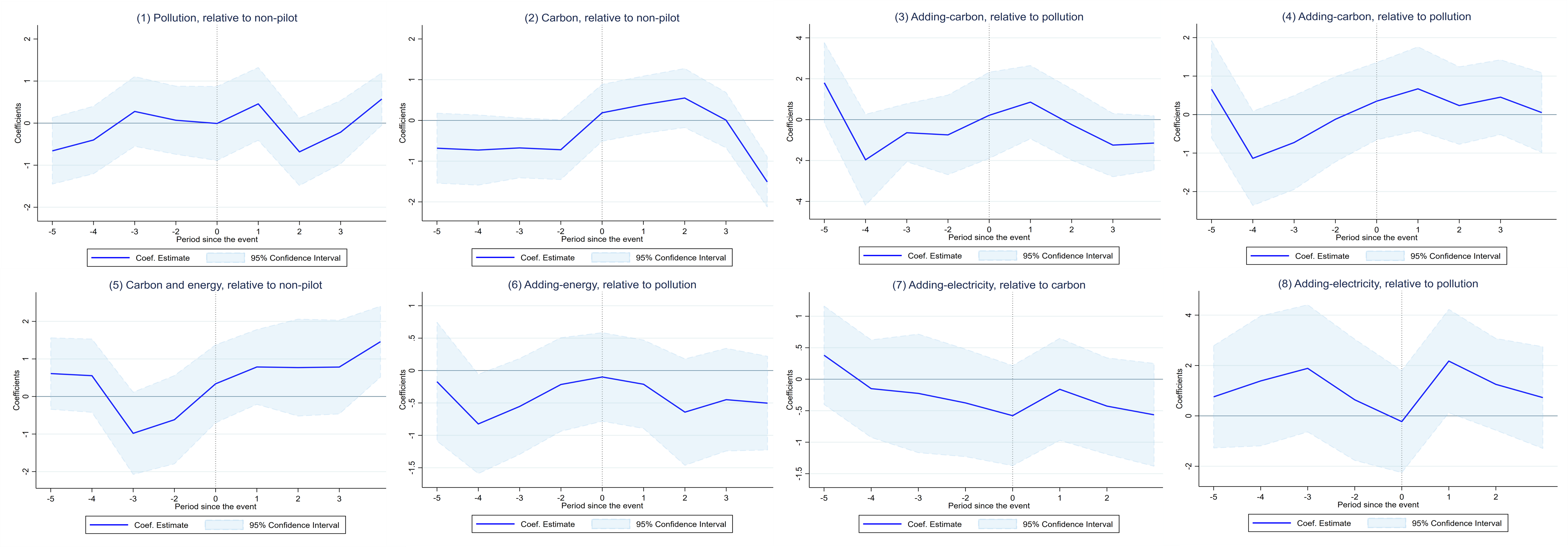}
    \caption{Event Study: Dynamic effects.}
    \label{fig:did_parallel_test}
\end{figure}

To ensure the validity of the DiD estimation, it is crucial that the treatment and control groups exhibit parallel trends in the outcome variable prior to the introduction of the carbon market. Violation of this assumption may lead to biased estimates and misinterpretation of the policy effect. We implement an event study by interacting year-specific dummy variables with the treatment group indicator, as follows:
\begin{align}
    ROA_{it} = \alpha + \sum_{j\neq -1,\, j\ge -4}^{4} \theta_j D_{i,t-j} + \delta X_{it} + \mu_i + \lambda_t + \varepsilon_{it}
\end{align}
where $ROA_{it}$ denotes the return on assets of company $i$ in year $t$, which serves as the dependent variable; $D_{i,t-j}$ is a set of dummy variables indicating the time distance $j$ from the treatment year for company $i$. The period $j = -1$ is omitted and serves as the baseline year; $\theta_j$ captures the effect of being $j$ years away from the policy implementation, relative to the policy year; $X_{it}$ is a vector of control variables; $\mu_i$ and $\lambda_t$ represent company fixed effects and year fixed effects, respectively; $\varepsilon_{it}$ is the error term. The coefficients $\theta_j$ and their 95\% confidence intervals are visualized in Figure~\ref{fig:did_parallel_test}. 
Figure~\ref{fig:did_parallel_test} reveals that Panels (4) and (7) exhibit a clear upward trend even before the policy implementation, implying that the corresponding DiD estimates may be biased. To address this concern, we present the ArCo results as a supplementary and comparative analysis.

\subsection{Results of ArCo}
\begin{figure}[H]
    \centering
    \includegraphics[width=\textwidth]{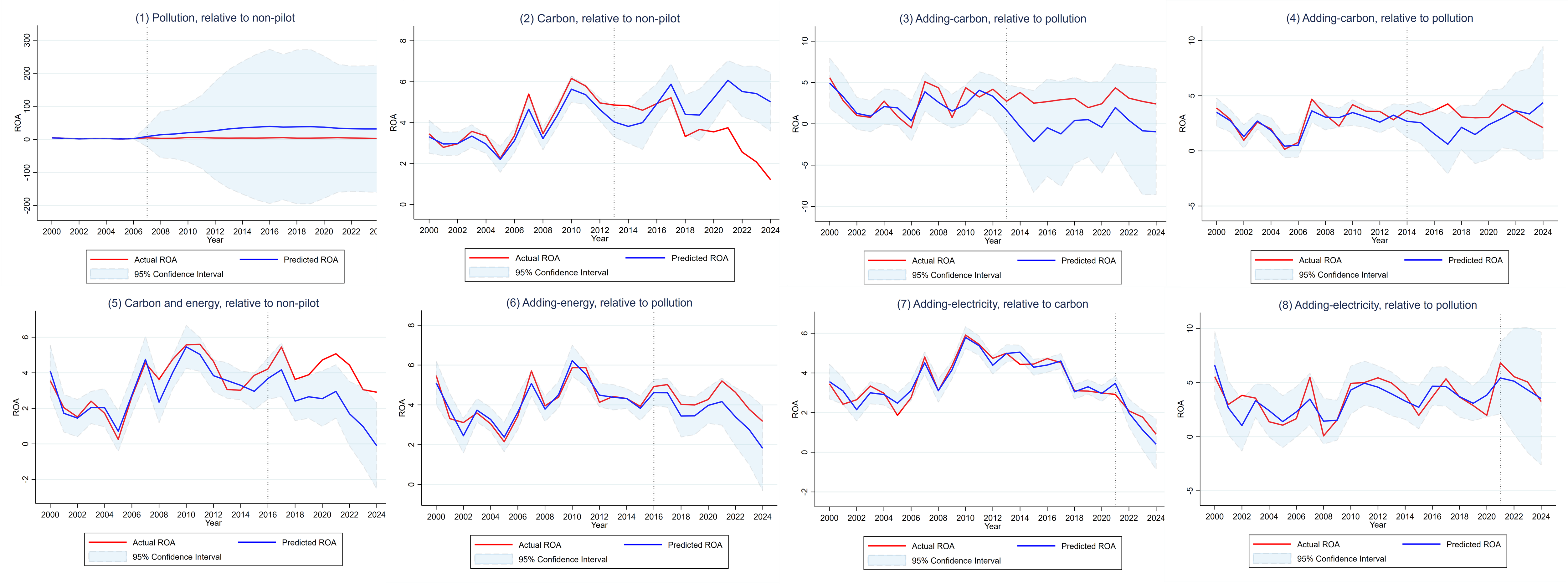}
    \caption{ArCo estimates for the eight panels.}
    \label{fig:ArCo}
\end{figure}

Table~\ref{tab:ArCo_main_results} presents parameters for pre-treatment fit of ArCo using the untreated companies as explanatory variables. Figure~\ref{fig:ArCo} presents the treatment effects estimated by ArCo of the eight panels.
\input{Tables.tex/Table-ArCo_main_results}

As shown in Figure~\ref{fig:ArCo}, the ArCo model in Panel (1) does not perform well in fitting and predicting the effects of pollution emission trading. The short pre-treatment period limits ArCo’s ability to fit the model accurately and make reliable predictions, as evidenced by the wide confidence intervals.

Panels (2) and (5) indicate that, relative to non-pilot regions, carbon emission trading tends to suppress companies’ ROA, whereas the simultaneous implementation of carbon emission trading and energy-use rights trading appears to enhance companies’ ROA. A comparison between Panel (2) and Panels (3) and (4) further shows that pollution emission trading mitigates the negative impact of carbon emission trading on companies’ ROA.

In addition, Panels (3), (4), (6), (7), and (8) consistently show that the introduction of energy-use rights trading or green electricity trading on top of existing carbon or pollution emission trading does not produce any further adding or marginal effects. These findings are consistent with the conclusions drawn from the DiD estimates.

The mechanisms underlying these findings are twofold. Compared to a single market mechanism, multiple overlapping environmental markets provide companies with opportunities for cross-market arbitrage. Additionally, companies with prior experience in emission trading may be better positioned to leverage such experience to optimize industrial restructuring and enhance resource allocation efficiency when participating in newly established markets.

\subsection{Interpretation}
Table~\ref{tab:did_arco_ATE} presents the average treatment effects estimated by DiD and ArCo of the eight panels.

\input{Tables.tex/Table-DiDArCo}

As previously discussed, the ArCo model in Panel (1) performs poorly in fitting and predicting the effects of pollution emission trading due to the short pre-treatment period. This limitation undermines the credibility of ArCo’s estimates in this case. Apart from this, the results of DiD and ArCo are generally consistent in terms of statistical significance. However, discrepancies arise between the two methods regarding the direction of the estimated coefficients in terms of the \textit{additional} effects. To explain this divergence, Table~\ref{tab:method_comparison} compares the underlying causal inference frameworks of the two approaches.

\input{Tables.tex/Table-method_comparison}

(1) Explanation of divergent estimates direction of \textit{additional} effects

The divergence in coefficient signs observed in the \textit{additional} effect models (e.g., Panel (3), (4), (6) and (7)) arises from the fundamental differences in the identification assumptions of the DiD and ArCo methods. Specifically, the DiD estimator relies on the parallel trends assumption to ensure causal identification, which requires that the treated and control groups exhibit similar outcome trajectories during the pre-treatment period. Formally, this assumption can be expressed as:
\begin{equation}
E[ROA_{1t}^{(0)} - ROA_{0t}^{(0)}] = \text{constant}, \quad \forall t < T_0
\end{equation}
Under this assumption, the DiD estimator computes the average treatment effect as the difference in post-treatment means between the treated and control groups:
\begin{equation}
\hat\beta = \frac{1}{T_1} \sum_{t = T_0}^{T} \left( ROA_{1t} - \bar{ROA}_{0t} \right)
\end{equation}
However, in the context of China’s environmental market reforms, companies involved in later-stage policies, such as energy-use rights and green electricity trading, often exhibit stronger pre-treatment growth due to prior investments in green transformation. DiD fails to account for this inherent upward trend, resulting in bias as it attributes post-treatment differences\textemdash partially driven by this natural momentum\textemdash to the policy itself.

In contrast, ArCo does not rely on the parallel trends assumption but constructs counterfactual trajectories through pre-treatment covariate-outcome modeling (cf. Table~\ref{tab:method_comparison}):
\begin{equation}
\hat{ROA}^{(0)}_{1t} = X_{1t}' \hat\theta
\end{equation}
The treatment effect at each period is then estimated as:
\begin{equation}
\hat\delta_t = ROA_{1t} - \hat{ROA}^{(0)}_{1t}
\end{equation}
By explicitly modeling the relationship between covariates and outcomes, ArCo is more robust to heterogeneous pre-treatment trends. This explains why ArCo often yields opposite signs compared to DiD in settings where treated companies exhibit strong pre-treatment growth trajectories. DiD is biased in such cases, but ArCo is not.

(2) Explanation of divergent statistical significance in Panel (1)

The difference in statistical significance observed in Panel (1) between DiD and ArCo primarily stems from the short pre-treatment period. ArCo requires a sufficiently long and stable pre-treatment window to reliably fit the counterfactual model. In Panel (1), where pollution trading is examined, the pre-treatment period is relatively short, leading to unreliable estimation of $\hat\theta$ and, consequently, low confidence in the construction of $\hat{ROA}^{(0)}_{1t}$. 

By contrast, DiD computes post-treatment differences in means and pools residuals to estimate standard errors globally:
\begin{equation}
\widehat{SE}(\hat\beta) = \sqrt{ \hat\sigma^2 \cdot (X'X)^{-1}_{jj} }, \quad \hat\sigma^2 = \frac{1}{n - k} \sum_{i,t} \hat\varepsilon_{it}^2
\end{equation}
Although this approach risks underestimating uncertainty in staggered treatment settings, it is less sensitive to short pre-treatment periods and therefore yields more stable, albeit potentially biased, estimates. This explains why DiD in Panel (1) reports statistically significant results, while ArCo does not.

\subsection{Robustness tests of DiD and ArCo estimates}

\subsubsection{The influence of industrial factors}
In the baseline analysis, a multi-layered counterfactual design is constructed by leveraging the staggered timing and spatial overlap of various environmental market pilots, in order to mitigate bias and threats to identification caused by policy overlaps or temporal misalignment. The robustness check introduces an additional industry-level restriction by retaining only those sectors simultaneously subject to carbon emission trading, carbon emission trading, energy-use rights trading, and green electricity trading. This restriction ensures that the treatment and control groups share a common policy exposure history prior to the implementation of the specific pilot under study\textemdash that is, both have previously been regulated under similar environmental markets\textemdash thereby enhancing group comparability. By excluding industries that had never been affected by a particular policy, this approach reduces structural heterogeneity and improves the explanatory validity of the estimated effects.

\begin{figure}[H]
    \centering
    \includegraphics[width=\textwidth]{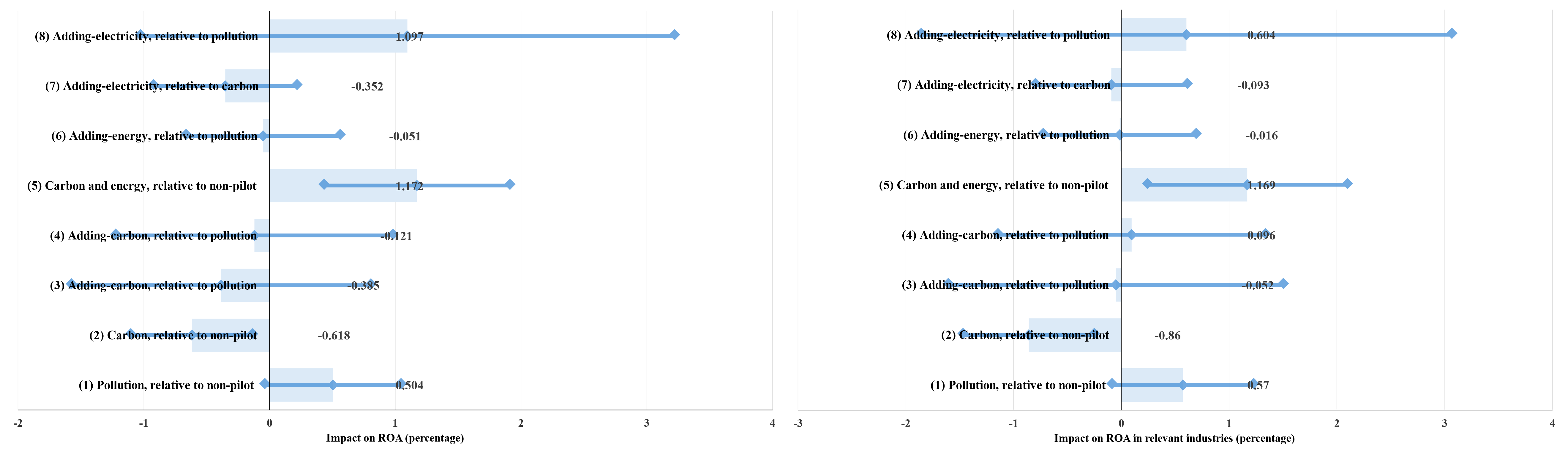}
    \caption{DiD estimates for ROA in relevant industries.}
    \raggedright
    \textit{Note: 95\% confidence intervals are shown in the figure; Panel (4) reports estimates based on CSDiD.}
    \label{fig:DiD-indus}
\end{figure}

\begin{figure}[H]
    \centering
    \includegraphics[width=\textwidth]{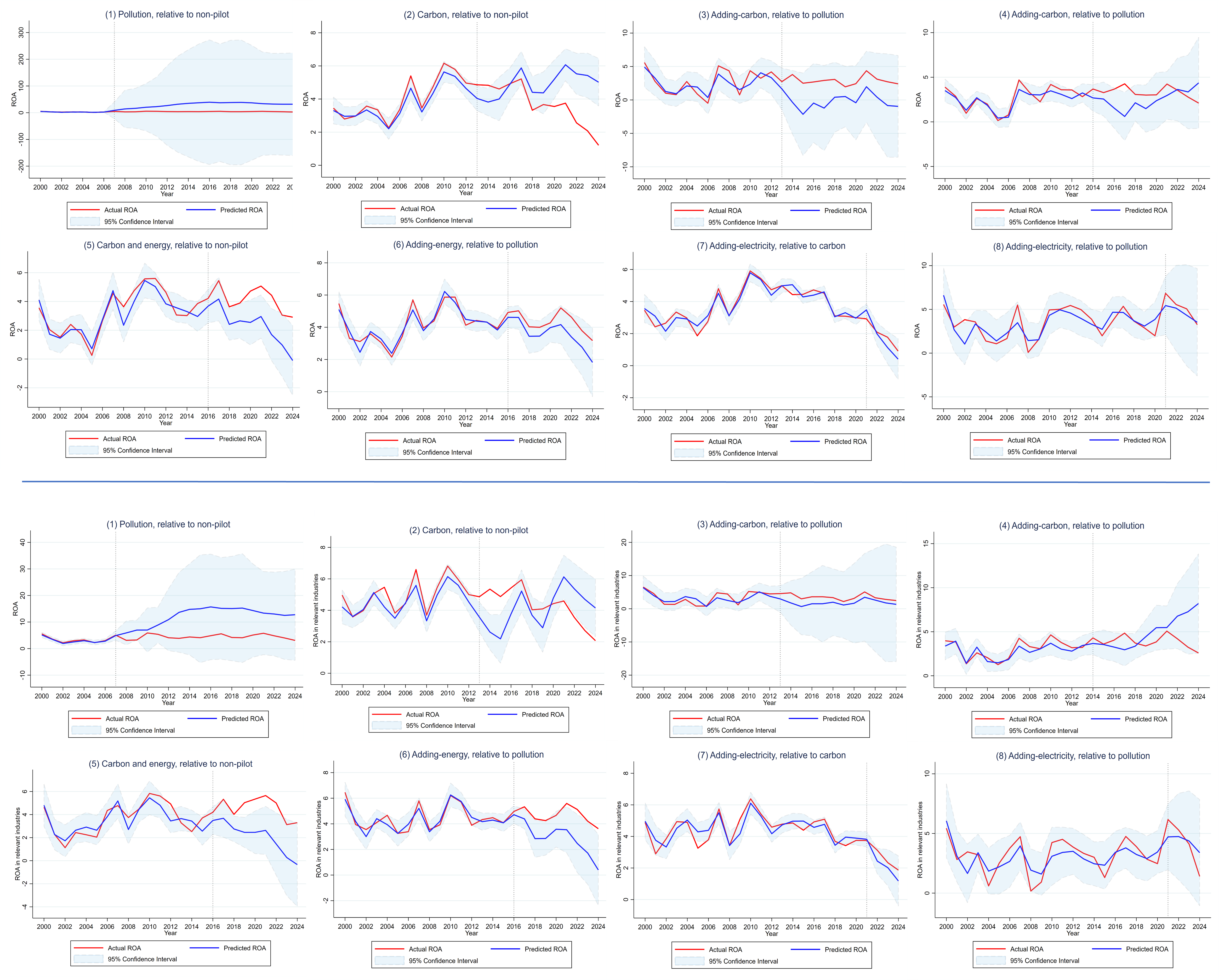}
    \caption{ArCo estimates for ROA in relevant industries.}
    \label{fig:ArCo-indus}
\end{figure}

Accordingly, we systematically compile the industry coverage of environmental market pilots based on official documents issued by local governments (Appendix \ref{Appendix-carbon indus}, \ref{Appendix-energy indus} and \ref{Appendix-electricity indus}), and map them to the industry categories used in the stock exchange classification system (Appendix \ref{Appendix-indus classification}), based on which the regression sample is further refined. We find that all types of environmental markets across different regions commonly cover industry codes C, D, and G. Accordingly, we retain only these three industries in the final sample and re-estimate the regressions across the eight panels. The results based on DiD and ArCo are presented in Figure~\ref{fig:DiD-indus} and Figure~\ref{fig:ArCo-indus}, respectively.

Figure~\ref{fig:DiD-indus} and Figure~\ref{fig:ArCo-indus} indicate that the DiD and ArCo estimates remain consistent in both sign and statistical significance. These findings confirm the robustness of the main regression results.

\subsubsection{The influence of tax}
\begin{figure}[H]
    \centering
    \includegraphics[width=\textwidth]{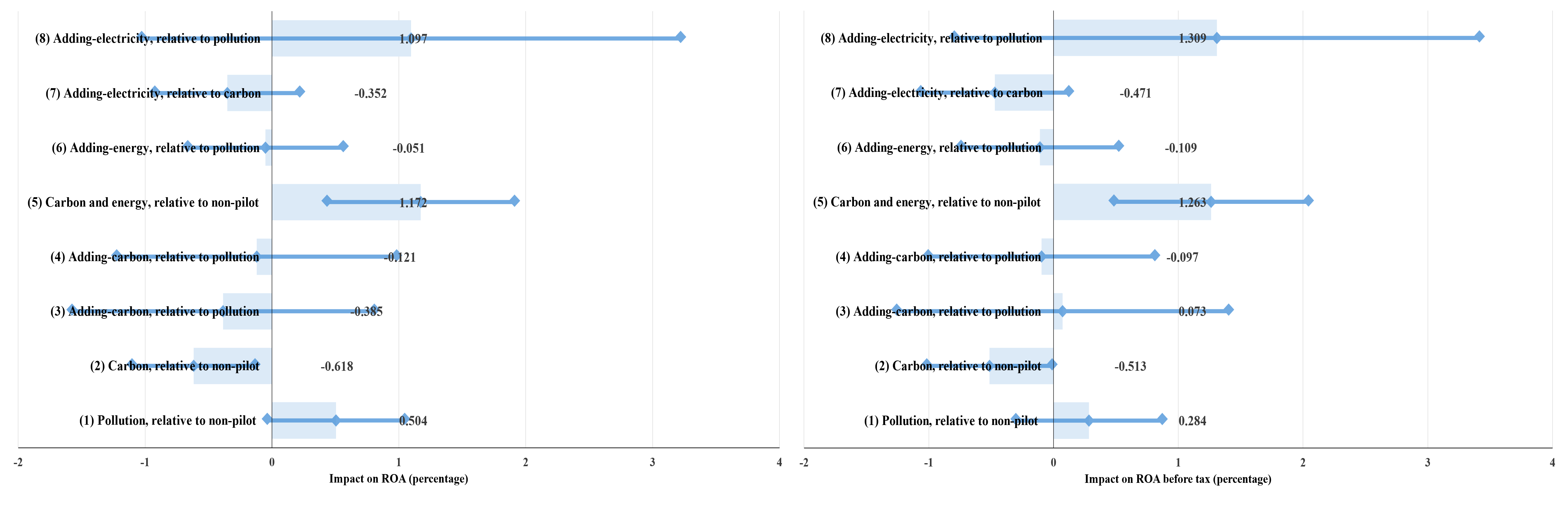}
    \caption{DiD estimates for ROA before tax.}
    \label{fig:DiD-tax}
    \raggedright
    \textit{Note: 95\% confidence intervals are shown in the figure; Panel (4) reports estimates based on CSDiD.}
\end{figure}

\begin{figure}[H]
    \centering
    \includegraphics[width=\textwidth]{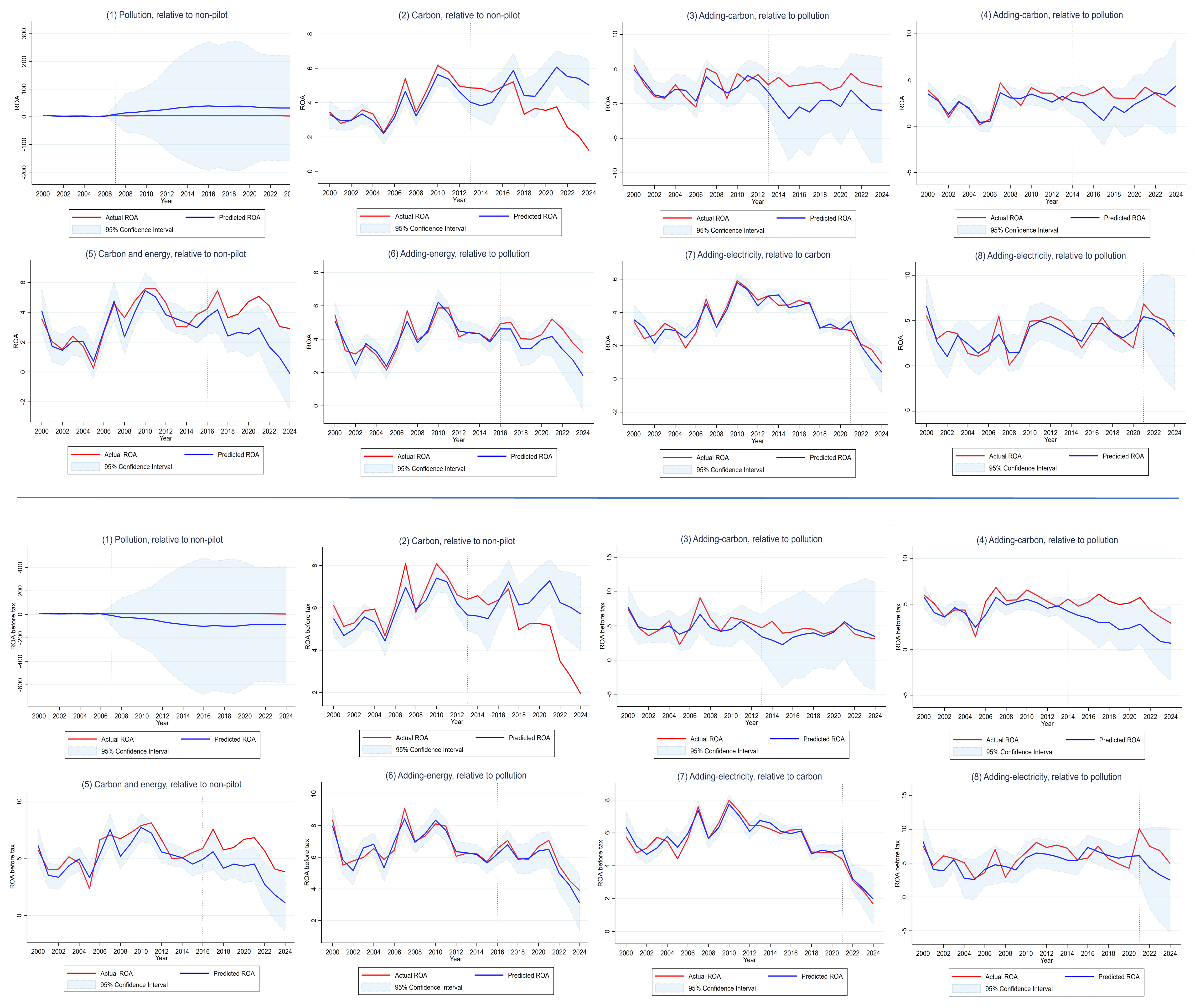}
    \caption{ArCo estimates for ROA before tax.}
    \label{fig:ArCo-tax}
\end{figure}

Policy effects may be mediated through fiscal mechanisms. For instance, companies participating in environmental markets often benefit from preferential tax treatments, such as exemptions, rebates, deductions, or direct fiscal subsidies. Consequently, the observed increase in post-tax ROA may partly reflect tax incentives rather than genuine improvements in operational performance. To address this concern, we recalculate the ROA before tax (see Table~\ref{tab:ROAtax}) to replace the original ROA. We then re-estimate the models for the nine sub-samples using the revised metric, and the DiD and ArCo results are presented in Figure~\ref{fig:DiD-tax} and Figure~\ref{fig:ArCo-tax}, respectively.

\begin{table}[H]
    \centering
    \caption{ROA before and after tax.}
    \scriptsize
    \begin{tabular}{lc}
    \toprule
    \textbf{Abatement cost} & \textbf{Description} \\
    \midrule
    ROA & Net profit/Average total assets \\
    ROA before tax & (Total profit + Financial expenses)/Average total assets \\
    \bottomrule
    \end{tabular}\\
    \raggedright
    \textit{Note: If the denominator is unavailable or equals zero, the result is recorded as NULL. Average Total Assets = (Ending Balance of Total Assets + Beginning Balance of Total Assets) / 2}
    \label{tab:ROAtax}
\end{table}

Figure~\ref{fig:DiD-tax} and Figure~\ref{fig:ArCo-tax} indicate that the regression outcomes based on pre-tax ROA exhibit a similar level of statistical significance as those based on post-tax ROA, regardless of whether ArCo or DiD is employed. This consistency reinforces the robustness of the baseline estimates.

\section{Discussion}
\label{sc:discussion}
\subsection{The overall influence of environmental markets on companies' return on assets}
While the primary contribution of this study lies in identifying the marginal effects of environmental markets, it is also necessary to revisit the overall impact of such markets on companies' abatement costs. This helps to align the present research with existing literature and provides a broader interpretation of the policy’s economic consequences.

Most studies on environmental markets adopt the TWFE DiD framework and focus primarily on the overall treatment effects of environmental markets. While these findings provide valuable insights into long-term policy outcomes, the TWFE DiD approach suffers from known biases arising from time-varying treatments \citep{callaway2021difference}, multiple treatments \citep{goldsmith2024contamination}, and potential interactions between pre-existing or concurrently implemented policies. These issues are particularly problematic in the Chinese context, where environmental markets have been implemented in multiple phases with staggered regional participation and overlapping policies.

\begin{figure}[H]
    \centering
    \includegraphics[width=\textwidth]{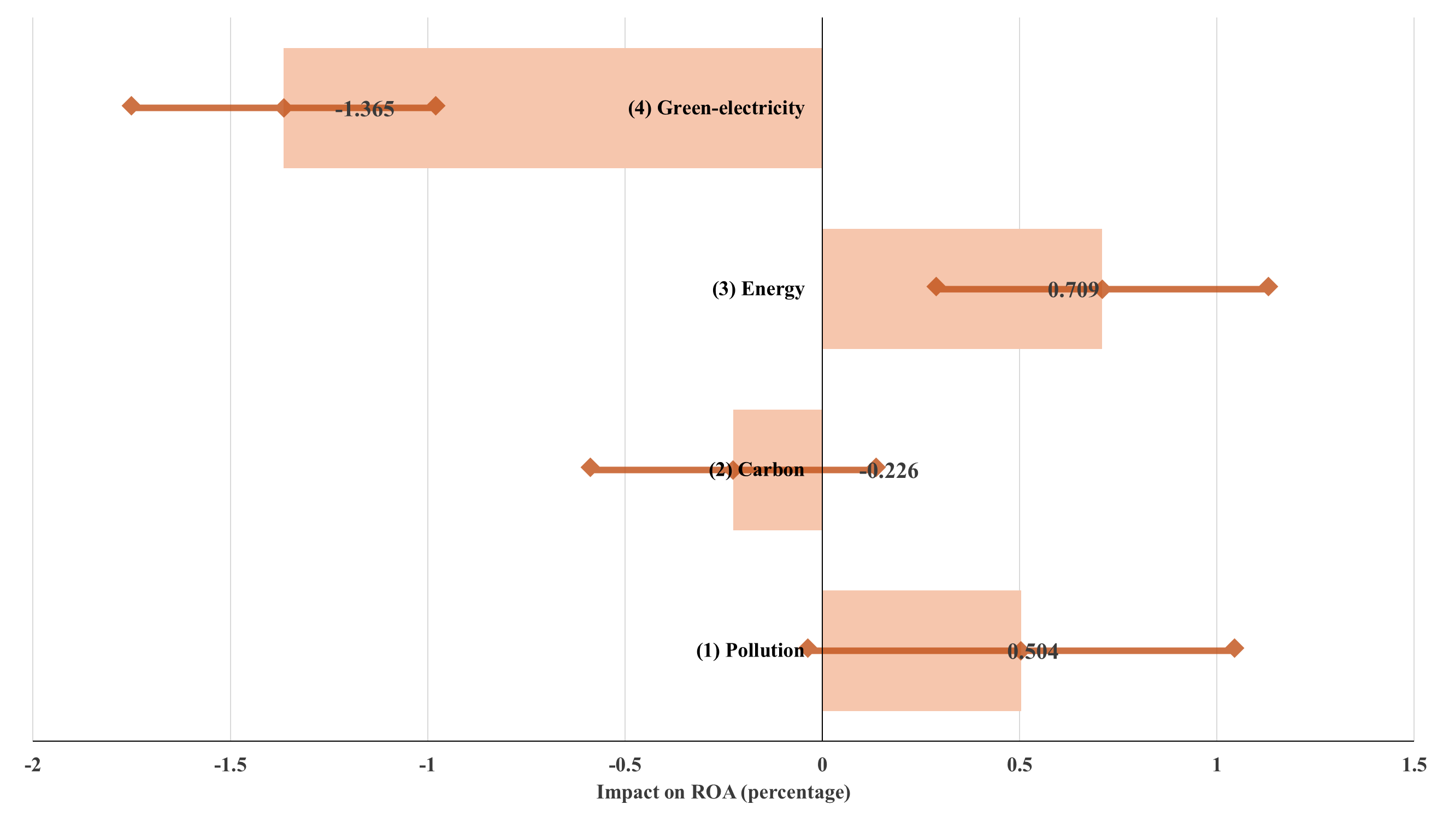}
    \caption{DiD estimates for ROA with static control group.}
    \label{fig:DiD-overall}
\end{figure}

To ensure comparability with prior research, we replicate the empirical strategies commonly employed in the literature, applying both DiD and ArCo methods. The TWFE DiD estimates are presented in Figure~\ref{fig:DiD-overall}.

Carbon emission trading pilots were introduced in different regions in 2013, 2014, and 2016. To account for this staggered implementation, we further employ the \textsc{csdid} approach proposed by \cite{callaway2021difference}, consistent with the methodology used in Panel (4). The results are reported in Table~\ref{tab:CSDID_overall}.

\input{Tables.tex/Table_CSDID_overall}

\begin{figure}[H]
    \centering
    \includegraphics[width=\textwidth]{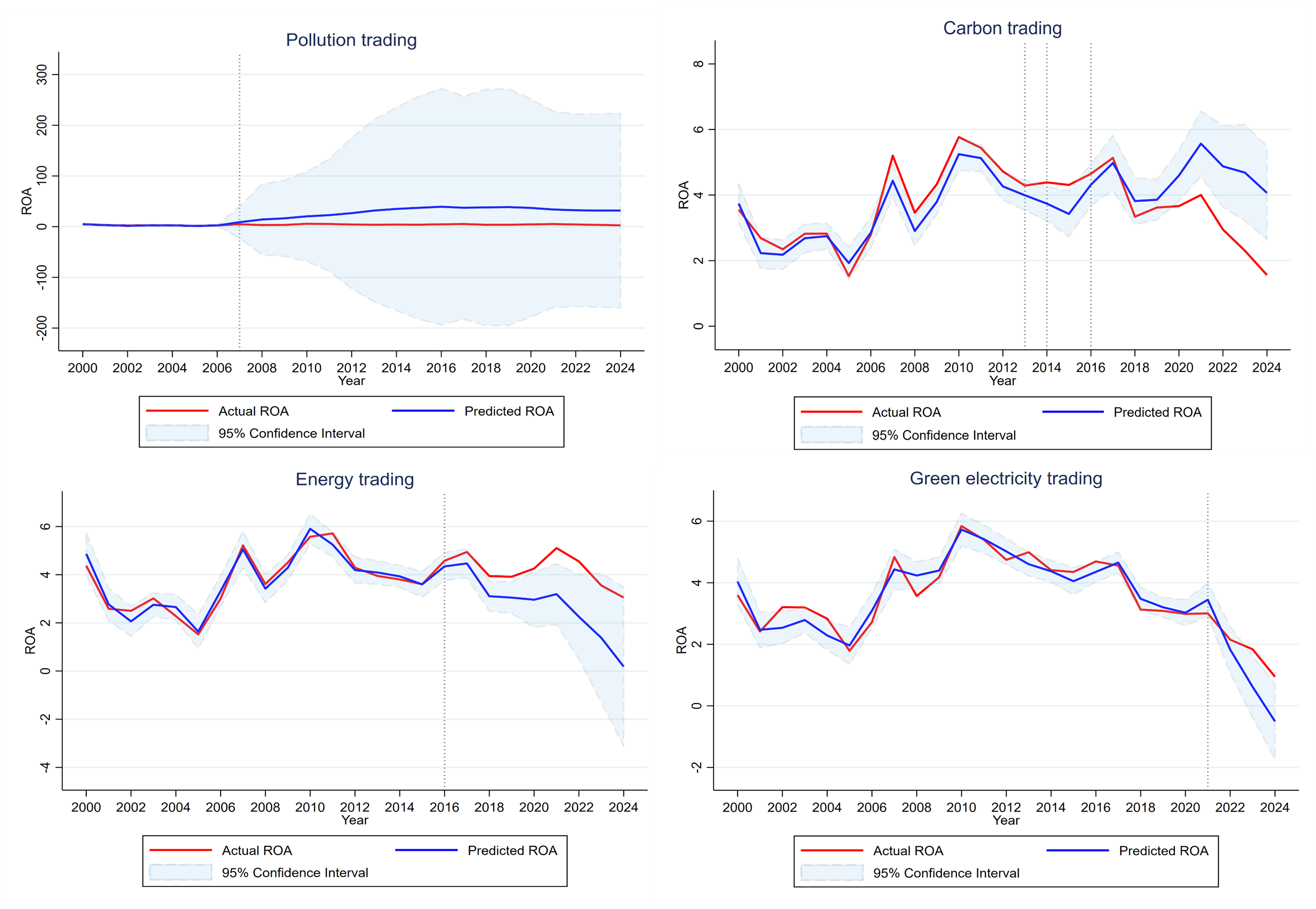}
    \caption{ArCo estimates for ROA with static control group.}
    \label{fig:ArCo-overall}
\end{figure}

The ArCo estimates are presented in Figure~\ref{fig:ArCo-overall}.

As shown in Figure~\ref{fig:DiD-overall} and Table~\ref{tab:CSDID_overall}, when using a static control group, the DiD regression results suggest that energy-use rights trading significantly improve companies’ ROA, whereas green electricity trading appears to suppress ROA, and carbon emissions trading is insignificant in TWFE staggered DiD but significant in staggered DiD improved by \cite{callaway2021difference}. The ArCo estimates in Figure~\ref{fig:ArCo-overall} align with the TWFE DiD results for energy-use rights trading and with the CSDiD estimates for carbon emission trading but diverge from the TWFE DiD estimates for carbon emission trading.

\subsection{Comparisons with existing literature}
Building on the replication results from the previous section, this part systematically compares three sets of estimates to assess how different empirical strategies influence the empirical conclusions. Table~\ref{tab:existing_study_comparison} summarizes the differences in coefficient signs across methods, including: (1) the original DiD estimates reported in the literature (columns labeled “Conclusions”); (2) our replication results using the same control group design, applying both DiD and ArCo (rows labeled “All”); and (3) our estimates after excluding contaminated samples (rows (1)–(8)).

\input{Tables.tex/Table_existing_study_comparison}

As shown in Table~\ref{tab:existing_study_comparison}, the existing literature presents mixed findings regarding the direction and magnitude of the impact of carbon emission trading on companies’ abatement costs. Our replication results align with the conclusions of \cite{li2025impact}, suggesting that carbon emission trading reduces companies’ ROA. However, after sequentially excluding contaminated samples, we find that the effect of carbon emission trading exhibits notable heterogeneity. Specifically, compared to companies in regions that had already implemented pollution emission trading, ROA shows no significant change following the introduction of carbon emission trading. In contrast, the negative effect becomes more pronounced and statistically significant when compared to companies in regions with no prior pilot programs.

For energy-use rights trading, the existing literature generally agrees that it promotes companies’ ROA, and our replication results are consistent with these findings. However, after excluding contaminated samples in a phased manner, we similarly observe heterogeneous effects. That is, relative to companies in regions already subject to pollution emission trading, energy-use rights trading does not lead to significant changes in ROA. In contrast, compared to companies in non-pilot regions, the positive effect becomes stronger and more statistically significant.

Previous studies commonly suggest that green electricity trading alleviates companies’ abatement costs, a conclusion supported by our replication results using the DiD approach. However, after employing cleaner samples, both the DiD and ArCo estimates reveal that this effect becomes statistically insignificant. This suggests that prior findings for the effects of green electricity trading may be biased, and such bias likely stems from the use of contaminated samples in previous studies.

\section{Conclusions}
\label{sc:conclusions}
This study examines the impact of China’s environmental markets on companies’ return on assets. To address the identification challenges posed by time-varying treatments, multiple overlapping policies, and their potential interactions with existing regulations, we apply phase-specific and region-specific DiD estimations by sequentially excluding contaminated samples. This allows us to capture the dynamic effects and marginal abatement costs associated with market implementation. Furthermore, we introduce the more flexible ArCo method to supplement and validate the DiD results. Both DiD and ArCo produce largely consistent conclusions regarding the significance and direction of policy effects, and these findings remain robust after accounting for industry and fiscal factors. Accordingly, we draw the following conclusions:

Carbon emissions trading reduces companies’ ROA, whereas existing pollution emissions trading mitigates this negative effect. Specifically, when comparing Panel (2) with Panels (3) and (4), the estimated effect changes from –0.618\% to a statistically insignificant negative value under the DiD approach, and from –1.044\% to a statistically insignificant positive value under the ArCo approach. Moreover, the simultaneous implementation of energy-use rights trading further offsets the adverse impact of carbon markets on firm performance. Comparing Panel (2) with Panel (5), the estimated effect shifts from -0.618\% to 1.172\% under DiD, and from –1.044\% to 1.750\% under ArCo, suggesting that overlapping environmental markets provide opportunities for cross-market arbitrage. 

Additionally, the further addition of energy-use rights or green electricity trading in regions already covered by carbon or pollution markets generates no significant marginal effects, as shown in Panel (6), (7), and (8) of Table~\ref{tab:identification}, indicating no additional financial costs or benefits from overlapping policies.

To ensure comparability with prior research, we replicate commonly used empirical strategies and find that the results are broadly consistent with prior literature, confirming the validity and comparability of our findings. However, after sequentially excluding contaminated samples, we observe clear heterogeneity. The effects of carbon emissions trading and energy-use rights trading on companies’ ROA are insignificant in regions already covered by pollution emissions trading (see Panels (3), (4), (6), (7) and (8) of Table~\ref{tab:identification}), but become more pronounced and statistically significant when compared to regions without prior pilot programs. Specifically, for carbon emissions trading (Panel (2)), the estimated effect changes from -0.404\% to -0.618\% under DiD and from -1.285\% to -1.944\% under ArCo. For energy-use rights trading (Panel (5)), the estimated effect shifts from 0.709\% to 1.172\% under DiD and from 1.635\% to 1.750\% under ArCo.

There are several caveats to our research. We study the impact of overlap between various permit markets, but local, provincial and national authorities use a range of additional policy instruments to affect (a) emissions and (b) profitability. We use province (and industry) to proxy ``treatment'' as we do not know which companies are actually regulated. CSMAR reports \emph{consolidated} accounts; we know the location and hence regulation of a firm's headquarters, but we do not know the location of its subsidiaries, let alone intra-firm reallocation in response to regulation. We observe listed companies, but unlisted ones are regulated too.

These caveats notwithstanding, we find that carbon permits reduced the return on assets, that energy-use permits increased the return on assets, while pollution permits and green electricity permits had no discernible effect.

\section*{Acknowledgment}
The authors are grateful for the financial support provided by China Scholarship Council.  

\newpage
\appendix
\renewcommand{\thetable}{A.\arabic{table}}
\setcounter{table}{0}

\section*{Appendix}

\input{Tables.tex/Appendix-carbon_indus}

\input{Tables.tex/Appendix-energy_indus}

\input{Tables.tex/Appendix-electricity_indus}

\input{Tables.tex/Appendix-indus_classification}

\clearpage
\bibliography{references}

%\appendix
%\setcounter{figure}{0}
%\renewcommand{\thefigure}{A\arabic{figure}}

\end{document}

%% file: Tables.tex/identification.tex
\begin{table}[]
    \centering
    \begin{tabular}{lllll}
    \hline
         Group & Year & Market & \# & Identification \\ \hline
         A0 & 2000-2006 & None & 31 & - \\ \hline
         A1 & 2007-2012 & None & 20 & - \\
         E1 & & Pollution & 11 & A1 v E1 (Panel 1): pollution  \\ \hline
         A2 & 2013 & None & 16 & - \\
         C2 & & CO\textsubscript{2} & 4 & A2 v C2 (Panel 2): CO\textsubscript{2} \\
         E2 & & Pollution & 10 & - \\
         F2 & & Poll + CO\textsubscript{2} & 1 & E2 v F2 (Panel 3): CO\textsubscript{2} conditional on pollution \\ \hline
         A3 & 2014-2015 & None & 16 & - \\
         E3 & & Pollution & 8 & - \\
         F3 & & Poll + CO\textsubscript{2} & 3 & E3 v F3 (Panel 4): CO\textsubscript{2} conditional on pollution \\ \hline
         A4 & 2016-2020 & None & 14 & - \\
         B4 & & CO\textsubscript{2} + energy & 2 & B4 v A4  (Panel 5): CO\textsubscript{2} + energy \\
         C4 & & CO\textsubscript{2} & 4 & - \\
         E4 & & Pollution & 6 & - \\
         F4 & & Poll + CO\textsubscript{2} & 3 & - \\
         G4 & & Poll + Energy & 2 & E4 v G4 (Panel 6): Energy conditional on pollution \\ \hline
         A5 & 2021-2024 & None & 14 & - \\
         B5 & & CO\textsubscript{2} + energy & 2 & - \\
         C5 & & CO\textsubscript{2} & 2 & - \\
         D5 & & CO\textsubscript{2} + Green & 2 &  C5 v D5 (Panel 7): Green conditional on CO\textsubscript{2}  \\
         E5 & & Pollution & 5 & - \\
         F5 & & Poll + CO\textsubscript{2} & 3 & - \\
         G5 & & Poll + Energy & 2 & -\\
         H5 & & Poll + Green & 1 & E5 v H5  (Panel 8): Green conditional on pollution \\ \hline
         \multicolumn{5}{l}{Market for energy use permits cannot be identified.} \\
         \multicolumn{5}{l}{Market for green electricity credits cannot be identified.} \\ \hline
    \end{tabular}
    \caption{Periods and provincial coverage of eight observed permutations of environmental markets and the implied identification.}
    \label{tab:identification}
\end{table}

%% file: Tables.tex/Table-data_description.tex
\begin{table}[htbp]
  \centering
  \caption{Key Variables and descriptions}
  \label{tab:variable_description}
  \begin{tabular}{@{}lp{9.5cm}@{}}
    \toprule
    Variable & Description \\
    \midrule
    ROA    & Return on Assets, defined as Net Profit divided by the Average Total Assets Balance. If asset balance is missing or zero, the value is coded as NULL. Average total assets = (Ending + Beginning total assets) / 2. Net Profit is taken from the consolidated income statement (including the parent company and all consolidated subsidiaries, net of minority interest), and Total Assets are from the consolidated balance sheet (including all consolidated subsidiaries).\\
    HHI    & Herfindahl-Hirschman Index, capturing market concentration at the industry level. \\
    AGE    & Firm age measured as the natural logarithm of years since establishment. \\
    EMP    & Total number of employees, in logarithmic form. \\
    OCCAR  & Operational Capital to Current Assets Ratio, capturing liquidity and capital allocation efficiency. \\
    DER    & Debt-to-Equity Ratio, indicating capital structure and financial leverage. \\
    \bottomrule
  \end{tabular}
\end{table}

%% file: Tables.tex/Table_data_statistics.tex
\begin{table}[htbp]
\centering
\caption{Summary statistics of key variables.}
\small
\renewcommand{\arraystretch}{1.2}
\begin{tabular}{llcccccc}
\toprule
Variable & Description & N & Mean & Median & SD & Min & Max \\
\midrule
ROA      & Return on assets &  61{,}993  & 3.69 & 3.72 & 6.88  & -26.2 & 22.2 \\
HHI      & Hirschman-Herfindahl index & 60{,}049  & 0.18 & 0.12 & 0.18  & 0.019 & 1 \\
lnAGE    & Age & 57{,}655  & 1.96 & 2.08 & 0.92  & 0     & 3.37 \\
lnEMP    & Number of employees & 61{,}871  & 7.57 & 7.51 & 1.30  & 4.19  & 11.2 \\
OCCAR    & Operational capital to current assets & 60{,}876  & 0.26 & 0.39 & 0.61  & -2.9  & 0.94 \\
DER      & Debt to equity & 61{,}994  & 1.27 & 0.73 & 1.87  & 0.023 & 13.1 \\
\bottomrule
\end{tabular}
\label{Table_data_statistics}
\end{table}

%% file: Tables.tex/Table-data_group_statistics.tex
\begin{table}[htbp]
\centering
\footnotesize
\caption{Summary statistics of ROA by the eight panels.}
\renewcommand{\arraystretch}{1.2}
\begin{tabular}{p{6.6cm} ccc ccc c}
\toprule 
 & \multicolumn{3}{c}{\textbf{Treatment}} & \multicolumn{3}{c}{\textbf{Control}} & \\
\cmidrule(lr){2-4} \cmidrule(lr){5-7}
 \textbf{Panels} & Obs & Mean & SD & Obs & Mean & SD & Difference \\
\midrule
(1) Pollution, relative to all & 20,982 & 4.158 & 6.644 & 41,011 & 3.443 & 6.979 & 0.715 \\
(2) Carbon, relative to no policy    & 14,351 & 3.397 & 6.977 & 24,008 & 3.551 & 7.025 & -0.154 \\
(3) Carbon, additional to pollution & 646 & 2.890 & 6.226 & 19,839 & 4.182 & 6.651 & -1.292 \\
(4) Carbon, additional to pollution & 2,353 & 3.260 & 6.828 & 18,132 & 4.256 & 6.609 & -0.996 \\
(5) Carbon and energy, relative to no policy & 2,476 & 4.096 & 7.549 & 35,883 & 3.452 & 6.967 & 0.644 \\
(6) Energy, additional to pollution & 5,474 & 4.657 & 6.821 & 11,929 & 4.128 & 6.514 & 0.529 \\
(7) Electricity, additional to carbon & 4,919 & 2.192 & 7.308 & 8,717 & 3.958 & 6.795 & -1.766 \\
(8) Electricity, additional to pollution & 105 & 5.172 & 7.763 & 9,701 & 3.908 & 6.626 & 1.264 \\
\bottomrule
\end{tabular}
\label{tab:data_group_statistics}
\end{table}

%% file: Tables.tex/Table-DiD_main_results.tex
\begin{table}[htbp]
\centering
\caption{DiD estimates for the eight panels.}
\scriptsize
\renewcommand{\arraystretch}{1.2}
\resizebox{\textwidth}{!}{
\begin{tabular}{lcccccccc}
\toprule
& (1) & (2) & (3) & (4) & (5) & (6) & (7) & (8) \\

& \multicolumn{1}{c}{Pollution,} 
& \multicolumn{1}{c}{Carbon,} 
& \multicolumn{1}{c}{Add. carbon,} 
& \multicolumn{1}{c}{Add. carbon,} 
& \multicolumn{1}{c}{Carbon \& energy,} 
& \multicolumn{1}{c}{Add. energy,}
& \multicolumn{1}{c}{Add. electricity,} 
& \multicolumn{1}{c}{Add. electricity,} \\

\multicolumn{1}{c}{relative to}
& \multicolumn{1}{c}{all} 
& \multicolumn{1}{c}{no policy} 
& \multicolumn{1}{c}{pollution} 
& \multicolumn{1}{c}{pollution} 
& \multicolumn{1}{c}{no policy} 
& \multicolumn{1}{c}{pollution}
& \multicolumn{1}{c}{carbon} 
& \multicolumn{1}{c}{pollution} \\

\cmidrule(lr){2-2} 
\cmidrule(lr){3-3} 
\cmidrule(lr){4-4} 
\cmidrule(lr){5-5} 
\cmidrule(lr){6-6}
\cmidrule(lr){7-7}
\cmidrule(lr){8-8}
\cmidrule(lr){9-9}

&\multicolumn{1}{c}{ROA}&\multicolumn{1}{c}{ROA}&\multicolumn{1}{c}{ROA}&\multicolumn{1}{c}{ROA}&\multicolumn{1}{c}{ROA}&\multicolumn{1}{c}{ROA}&\multicolumn{1}{c}{ROA}&\multicolumn{1}{c}{ROA}\\
\midrule
Treat\texttimes Post       &    0.504\sym{*}  &   -0.618\sym{**} &   -0.385         &    0.018         &    1.172\sym{***}&   -0.051         &   -0.352         &    1.097         \\
          &  (0.276)         &  (0.248)         &  (0.607)         &  (0.423)         &  (0.377)         &  (0.312)         &  (0.291)         &  (1.083)         \\
HHI      &   -0.141         &    0.356         &   -0.032         &   -0.018         &    0.408         &   -0.488         &    1.683         &   -1.737         \\
          &  (0.462)         &  (0.574)         &  (0.782)         &  (0.781)         &  (0.574)         &  (0.859)         &  (1.042)         &  (1.171)         \\
lnAGE     &   -1.464\sym{***}&   -1.427\sym{***}&   -1.633\sym{***}&   -1.630\sym{***}&   -1.427\sym{***}&   -1.717\sym{***}&   -1.976\sym{***}&   -1.679\sym{***}\\
          &  (0.093)         &  (0.119)         &  (0.161)         &  (0.161)         &  (0.119)         &  (0.172)         &  (0.200)         &  (0.237)         \\
lnEMP&    0.354\sym{***}&    0.336\sym{***}&    0.556\sym{***}&    0.556\sym{***}&    0.333\sym{***}&    0.778\sym{***}&    1.200\sym{***}&    0.750\sym{***}\\
          &  (0.080)         &  (0.097)         &  (0.175)         &  (0.175)         &  (0.097)         &  (0.181)         &  (0.212)         &  (0.232)         \\
OCCAR&    3.587\sym{***}&    3.585\sym{***}&    3.780\sym{***}&    3.779\sym{***}&    3.576\sym{***}&    3.759\sym{***}&    4.611\sym{***}&    3.402\sym{***}\\
          &  (0.157)         &  (0.196)         &  (0.314)         &  (0.314)         &  (0.197)         &  (0.343)         &  (0.388)         &  (0.422)         \\
DER&   -0.534\sym{***}&   -0.520\sym{***}&   -0.505\sym{***}&   -0.506\sym{***}&   -0.518\sym{***}&   -0.547\sym{***}&   -0.502\sym{***}&   -0.504\sym{***}\\
          &  (0.041)         &  (0.051)         &  (0.073)         &  (0.073)         &  (0.052)         &  (0.088)         &  (0.103)         &  (0.100)         \\
Constant    &    3.226\sym{***}&    3.558\sym{***}&    2.226\sym{*}  &    2.204\sym{*}  &    3.260\sym{***}&    0.794         &   -3.522\sym{**} &    1.011         \\
          &  (0.638)         &  (0.779)         &  (1.329)         &  (1.331)         &  (0.776)         &  (1.382)         &  (1.629)         &  (1.792)         \\
\midrule
N         &54219         &33665         &17912         &17912         &33665         &15154         &11935         & 8498         \\
R\textsuperscript{2}        &    0.450         &    0.445         &    0.493         &    0.493         &    0.445         &    0.504         &    0.542         &    0.497         \\
\bottomrule
\end{tabular}
}
\raggedright
\textit{Note: Standard errors clustered at the company level are reported in parentheses. * $p<0.1$, ** $p<0.05$, *** $p<0.01$}
\label{tab:DiD_main_results}
\end{table}

%% file: Tables.tex/Table_CSDID.tex
\begin{table}[htbp]
\centering
\caption{CSDiD estimates (ATT) for carbon on top of pollution trading.}
\scriptsize
\renewcommand{\arraystretch}{1.2}
\resizebox{\textwidth}{!}{
\begin{tabular}{lcccc}
\toprule
& \multicolumn{1}{c}{Simple Weighted} 
& \multicolumn{1}{c}{Before and after treatment} 
& \multicolumn{1}{c}{By group} 
& \multicolumn{1}{c}{By calendar period} \\

\midrule
Simple & -0.121 & & & \\
            & (-0.563) & & & \\
Avg before    &        & -0.723{**} & & \\
            &        & (-0.299) & & \\
Avg after  &        & -0.147 & & \\
            &        &  (0.579) & & \\
Group    &        &        & -0.092 & \\
            &        &        &  (0.558) & \\
Calendar    &        &        &        & -0.207 \\
            &        &        &        &  (0.571) \\
\bottomrule
\end{tabular}
}
\raggedright
\textit{Note: Standard errors are reported in parentheses. * $p<0.1$, ** $p<0.05$, *** $p<0.01$}
\label{tab:CSDID}
\end{table}

%% file: Tables.tex/Table-ArCo_main_results.tex
\begin{table}[htbp]
\centering
\caption{ArCo estimates for the eight panels.}
\scriptsize
\resizebox{\textwidth}{!}{%
\begin{tabular}{lccccccccc}
\toprule
& \multicolumn{1}{c}{(1)} 
& \multicolumn{1}{c}{(2)} 
& \multicolumn{1}{c}{(3)} 
& \multicolumn{1}{c}{(4)} 
& \multicolumn{1}{c}{(5)} 
& \multicolumn{1}{c}{(6)}
& \multicolumn{1}{c}{(7)} 
& \multicolumn{1}{c}{(8)} \\

& \multicolumn{1}{c}{Pollution,} 
& \multicolumn{1}{c}{Carbon,} 
& \multicolumn{1}{c}{Add. carbon,} 
& \multicolumn{1}{c}{Add. carbon,} 
& \multicolumn{1}{c}{Carbon \& energy,} 
& \multicolumn{1}{c}{Add. energy,}
& \multicolumn{1}{c}{Add. electricity,} 
& \multicolumn{1}{c}{Add. electricity,} \\

\multicolumn{1}{c}{relative to}
& \multicolumn{1}{c}{all} 
& \multicolumn{1}{c}{no policy} 
& \multicolumn{1}{c}{pollution} 
& \multicolumn{1}{c}{pollution} 
& \multicolumn{1}{c}{no policy} 
& \multicolumn{1}{c}{pollution}
& \multicolumn{1}{c}{carbon} 
& \multicolumn{1}{c}{pollution} \\

\cmidrule(lr){2-2} 
\cmidrule(lr){3-3} 
\cmidrule(lr){4-4} 
\cmidrule(lr){5-5} 
\cmidrule(lr){6-6}
\cmidrule(lr){7-7}
\cmidrule(lr){8-8}
\cmidrule(lr){9-9}

&\multicolumn{1}{c}{ROA}&\multicolumn{1}{c}{ROA}&\multicolumn{1}{c}{ROA}&\multicolumn{1}{c}{ROA}&\multicolumn{1}{c}{ROA}&\multicolumn{1}{c}{ROA}&\multicolumn{1}{c}{ROA}&\multicolumn{1}{c}{ROA}\\
\hline
HHI      &   -1.710\sym{***}&   -0.747\sym{*}  &    2.809\sym{**} &    1.044         &   -0.888         &   -1.757\sym{***}&   -0.255         &    2.881\sym{*}  \\
          &  (0.515)         &  (0.409)         &  (1.376)         &  (0.753)         &  (0.851)         &  (0.533)         &  (0.362)         &  (1.686)         \\
lnAGE     &   -1.339\sym{***}&   -0.584\sym{***}&    0.591         &   -0.306         &   -0.135         &   -0.527\sym{***}&   -0.832\sym{***}&   -1.670\sym{***}\\
          &  (0.196)         &  (0.112)         &  (0.548)         &  (0.244)         &  (0.213)         &  (0.135)         &  (0.081)         &  (0.531)         \\
lnEMP&    0.851\sym{***}&    0.481\sym{***}&    0.657\sym{**} &    0.564\sym{***}&    0.459\sym{***}&    0.889\sym{***}&    0.812\sym{***}&    1.062\sym{***}\\
          &  (0.111)         &  (0.058)         &  (0.262)         &  (0.122)         &  (0.140)         &  (0.094)         &  (0.049)         &  (0.212)         \\
OCCAR&    2.945\sym{***}&    2.016\sym{***}&    5.972\sym{***}&    3.026\sym{***}&    2.154\sym{***}&    3.554\sym{***}&    2.522\sym{***}&    1.248\sym{***}\\
          &  (0.221)         &  (0.130)         &  (0.706)         &  (0.255)         &  (0.236)         &  (0.233)         &  (0.124)         &  (0.398)         \\
DER&   -1.079\sym{***}&   -0.869\sym{***}&   -0.356\sym{*}  &   -0.701\sym{***}&   -0.840\sym{***}&   -0.872\sym{***}&   -0.894\sym{***}&   -0.590\sym{**} \\
          &  (0.098)         &  (0.064)         &  (0.203)         &  (0.100)         &  (0.083)         &  (0.089)         &  (0.051)         &  (0.295)         \\
\textoverline{ROA}\textsubscript{control}&   -0.674         &    0.787\sym{***}&    1.164\sym{**} &   -0.480         &    1.194\sym{***}&    0.849\sym{***}&    1.027\sym{***}&    1.449\sym{***}\\
          &  (4.185)         &  (0.204)         &  (0.478)         &  (0.573)         &  (0.293)         &  (0.190)         &  (0.146)         &  (0.532)         \\
\textoverline{HHI}\textsubscript{control}&    8.757         &  -20.641\sym{*}  &   -5.629         &   24.340\sym{**} &   10.716         &   18.672\sym{*}  &    9.257\sym{**} &  -32.742         \\
          &(120.708)         & (11.264)         & (33.008)         & (11.700)         & (15.799)         & (10.105)         &  (4.030)         & (25.774)         \\
\textoverline{AGE}\textsubscript{control}&   13.749         &    2.485\sym{*}  &  -11.512\sym{**} &   -0.015         &   -0.109         &    1.371         &    1.767\sym{**} &   -2.884         \\
          & (33.445)         &  (1.268)         &  (4.704)         &  (3.136)         &  (2.110)         &  (1.084)         &  (0.756)         &  (2.728)         \\
\textoverline{EMP}\textsubscript{control}&   60.225         &  -12.973\sym{***}&    1.662         &   -0.739         &    1.762         &    0.779         &    0.110         &   -3.808         \\
          &(207.845)         &  (3.431)         & (10.561)         &  (2.893)         &  (3.971)         &  (3.723)         &  (1.330)         & (10.741)         \\
\textoverline{OCCAR}\textsubscript{control}&   32.584         &    1.492         &    1.798         &   15.719\sym{**} &   -4.493         &   -3.660         &   -0.358         &    2.214         \\
          & (72.526)         &  (2.109)         &  (8.335)         &  (7.604)         &  (4.186)         &  (3.962)         &  (1.369)         &  (8.899)         \\
\textoverline{DER}\textsubscript{control}&    0.924         &    1.033         &   18.183\sym{*}  &   12.192\sym{**} &   -0.100         &    0.328         &    2.849\sym{*}  &    6.592         \\
          & (22.265)         &  (2.462)         &  (9.550)         &  (4.768)         &  (3.450)         &  (2.094)         &  (1.643)         &  (6.164)         \\
Constant    & -468.344         &   95.796\sym{***}&  -24.038         &  -18.248         &  -17.742         &  -16.027         &  -14.075         &   25.547         \\
          &(1517.878)         & (26.395)         & (83.580)         & (16.304)         & (29.644)         & (27.511)         &  (9.895)         & (87.572)         \\
\hline
N         & 2229         & 4593         &  300         & 1318         & 1596         & 2541         & 8348         &  400         \\
R\textsuperscript{2}        &    0.236         &    0.159         &    0.278         &    0.213         &    0.190         &    0.248         &    0.165         &    0.161         \\
\bottomrule
\end{tabular}%
}
\raggedright
\textit{Note: Standard errors in parentheses. * $p<0.1$, ** $p<0.05$, *** $p<0.01$}
\label{tab:ArCo_main_results}
\end{table}

%% file: Tables.tex/Table-DiDArCo.tex
\begin{table}[H]
\centering
\begin{threeparttable}
\caption{Comparison of DiD and ArCo estimates.}
\begin{tabular}{llcc}
\toprule
\textbf{Panels} & \textbf{Control group} & \textbf{DiD} & \textbf{ArCo} \\
\midrule
\multirow{2}{*}{(1) Pollution}          & \multirow{2}{*}{all} & 0.504$^*$     & -28.422     \\
                                       &                             & (0.276)       & (98.395)    \\
\multirow{2}{*}{(2) Carbon}            & \multirow{2}{*}{no policy} & -0.618$^{**}$& -1.944$^{***}$ \\
                                       &                             & (0.248)       & (0.596)     \\
\multirow{2}{*}{(3) Additional carbon}     & \multirow{2}{*}{pollution} & -0.385        & 2.590       \\
                                       &                             & (0.607)       & (3.098)     \\
\multirow{2}{*}{(4) Additional carbon}     & \multirow{2}{*}{pollution} & -0.121        & 0.223       \\
                                       &                             & (0.563)       & (1.660)     \\
\multirow{2}{*}{(5) Carbon and energy} & \multirow{2}{*}{no policy} & 1.172$^{***}$  & 1.750$^{**}$   \\
                                       &                             & (0.377)       & (0.888)     \\
\multirow{2}{*}{(6) Additional energy}     & \multirow{2}{*}{pollution} & -0.051        & 0.844       \\
                                       &                             & (0.312)       & (0.720)     \\
\multirow{2}{*}{(7) Additional electricity}& \multirow{2}{*}{carbon}    & -0.352        & 0.216       \\
                                       &                             & (0.291)       & (0.479)     \\
\multirow{2}{*}{(8) Additional electricity}& \multirow{2}{*}{pollution} & 1.097         & 0.193      \\
                                       &                             & (1.083)       & (2.642)     \\
\bottomrule
\end{tabular}
\raggedright
\textit{Note: Standard errors are reported in parentheses. Panel (4) reports estimates based on CSDiD. * $p<0.1$, ** $p<0.05$, *** $p<0.01$.} 
\label{tab:did_arco_ATE}
\end{threeparttable}
\end{table}

%% file: Tables.tex/Table-method_comparison.tex
\begin{table}[H]
\centering
\footnotesize
\caption{Comparison of inference components between DiD and ArCo.}
\begin{tabular}{
  >{\raggedright\arraybackslash}p{3.5cm} 
  >{\raggedright\arraybackslash}p{5.2cm} 
  >{\raggedright\arraybackslash}p{5.2cm} 
}
\toprule
 & \textbf{DiD} & \textbf{ArCo} \\
\midrule

\textbf{Residuals}
& Regression residuals: $\hat{\varepsilon}_{it} = Y_{it} - X_{it}'\hat{\beta}$ \par Error variance: $\hat{\sigma}^2 = \frac{1}{n - k} \sum \hat{\varepsilon}_{it}^2$
& Residuals from untreatment regression: $\hat{\varepsilon}_{it} = Y_{it} - X_{it}'\hat{\theta}$ \par Error variance: $\hat{\sigma}^2 = \frac{1}{n_c - k} \sum_{i \in C} \hat{\varepsilon}_{it}^2$ \\

\midrule

\textbf{Standard error}
& $SE(\hat{\beta}_j) = \sqrt{[\widehat{\mathrm{Var}}(\hat{\beta})]_{jj}}$, where $\widehat{\mathrm{Var}}(\hat{\beta})] = \hat{\sigma}^2 [(X'X)^{-1}]_{jj} $ (without cluster) 
\par  Measures uncertainty of coefficient $\beta$
& $SE_t = \sqrt{X_t' \widehat{\mathrm{Var}}(\hat{\theta}) X_t}$, where $\widehat{\mathrm{Var}}(\hat{\theta}) = \hat{\sigma}^2 (X'X)^{-1}$ 
\par  Measures prediction uncertainty of the counterfactual $\hat Y_{1t}^{(0)}$
\\

\midrule

\textbf{Confidence interval}
& $\hat{\beta} \pm z_{1 - \alpha/2} \cdot SE(\hat{\beta})$ \par An average treatment effect CI
& $\hat{Y}_{1t}^{(0)} \pm z_{1 - \alpha/2} \cdot SE_t$ \par A time-varying CI at the predicted counterfactual\\

\midrule

\textbf{Target of inference}
& $H_0: \beta = 0$ 
& $H_0: \delta_t = Y_{1t} - \hat{Y}^{(0)}_{1t} = 0$ \\

\bottomrule
\end{tabular}
\label{tab:method_comparison}
\end{table}

%% file: Tables.tex/Table_CSDID_overall.tex
\begin{table}[htbp]
\centering
\caption{CSDiD estimates (ATT) for carbon dioxide emission trading.}
\scriptsize
\renewcommand{\arraystretch}{1.2}
\resizebox{\textwidth}{!}{
\begin{tabular}{lcccc}
\toprule
& \multicolumn{1}{c}{Simple Weighted} 
& \multicolumn{1}{c}{Before and after treatment} 
& \multicolumn{1}{c}{By group} 
& \multicolumn{1}{c}{By calendar period} \\

\midrule
Simple ATT  & -0.404{*} & & & \\
            & (0.218) & & & \\
Avg before    &        & 0.202 & & \\
            &        & (0.130) & & \\
Avg after   &        & -0.478{**} & & \\
            &        &  (0.220) & & \\
Group    &        &        & -0.376{*} & \\
            &        &        &  (0.219) & \\
Calender    &        &        &      &-0.366{*} \\
            &        &        &      &  (0.214) \\
\bottomrule
\end{tabular}
}
\raggedright
\textit{Note: Standard errors are reported in parentheses. * $p<0.1$, ** $p<0.05$, *** $p<0.01$}
\label{tab:CSDID_overall}
\end{table}

%% file: Tables.tex/Table_existing_study_comparison.tex
\begin{table}[htbp]
\centering
\caption{Comparison with existing literature and replication.}
\resizebox{\textwidth}{!}{%
\begin{tabular}{p{3.6cm}llp{2.2cm}p{7.8cm}lll}
\toprule
 & \multicolumn{4}{c}{\textbf{Existing studies}} & \multicolumn{3}{c}{\textbf{This paper}} \\
\cmidrule(lr){2-5} \cmidrule(lr){6-8}
\textbf{Markets} & \textbf{Control} & \textbf{Study} & \textbf{Methods} & \textbf{Conclusions} & \textbf{Control} & \textbf{DiD} & \textbf{ArCo} \\
\midrule
\textbf{Pollution emission trading in 2007} 
    & All & \cite{chen2022environmental} & DiD 
    & The emissions trading program is negatively associated with real earnings management. 
    & (1) All & 0.504{*} & -28.422 \\
    & & \cite{liu2022so2} & DiD & Our findings support the strong version of the Porter hypothesis.\\
\midrule
\textbf{Carbon emission trading in 2013 \& 2014} 
    & All & \cite{luan2025impact} & DiD
    & Regulated enterprises exhibit significantly better average economic performance 
    & (2) Non-pilot & -0.618{**} & -1.944{***} \\
 & & & & & (3) Pollution & -0.385 & 2.590 \\
 & & \cite{li2025impact}& DiD
    & Carbon emission trading system significantly increases the implied cost of equity capital for firms in the pilot areas. 
    & (4) Pollution & -0.121 & 0.223 \\
\cmidrule(lr){6-8}
 & & & & & All & -0.226/-0.404{*} & -1.285{**} \\
\midrule
\textbf{Energy-use rights and carbon emission trading in 2016} 
    & All & \cite{wang2024impact}& DiD 
    & China's energy-consuming rights trading can alleviate firms' financial resource mismatch. 
    & (5) Non-pilot & 1.172{***} & 1.750{***} \\
 & &\cite{wang2025towards}& DiD 
    & The energy right trading policy is helpful to improve the carbon performance. 
    & (6) Pollution & -0.051 & 0.844 \\
\cmidrule(lr){6-8}
 & & & & & All & 0.709{***} & 1.635{*} \\
\midrule
\textbf{Green electricity trading in 2021} 
    & All & \cite{tang2023does}& DiD 
    & Green power trading significantly alleviates the policy-covered firms' debt burden. 
    & (7) Carbon & -0.352 & 0.216 \\
 & & & & & (8) Pollution & 1.097 & 0.193 \\
\cmidrule(lr){6-8}
 & & & & & All & -1.365{***} & 0.7091 \\
\bottomrule
\end{tabular}
}
\vspace{0.5em}
\raggedright
\textit{Note: * $p<0.1$, ** $p<0.05$, *** $p<0.01$}
\label{tab:existing_study_comparison}
\end{table}

%% file: Tables.tex/Appendix-carbon_indus.tex
\begin{table}[htbp]
\centering
\caption{Carbon emission trading pilots and coverage.}
\scriptsize
\renewcommand{\arraystretch}{1.4}
\begin{tabular}{p{2.3cm}p{1.2cm}p{9.8cm}p{1.8cm}}
\toprule
\textbf{Pilots} & \textbf{Start year} & \textbf{Coverage scope (summarized from official documents)} & \textbf{Industry classification codes} \\
\midrule
Shenzhen & 2013 & Power supply, water supply, gas supply; data centers; public transport; metro systems; hazardous waste treatment, solid waste, sludge, and wastewater treatment; ports and terminals; flat panel display, information-based chemicals and other specialty chemicals; manufacturing and other sectors. & C, D, G, I, N \\
Guangdong & 2013 & Power generation, cement, steel, petrochemicals, papermaking, civil aviation, ceramics, construction, sanitation, and transportation. & C, D, G, E \\
Shanghai & 2013 & Power generation, power grid, and heat supply industries; auto glass production; data centers; steel, petrochemicals, chemicals, non-ferrous metals, building materials, textiles, papermaking, rubber, chemical fibers and other industrial enterprises; aviation, ports, shipping, tap water supply enterprises; shopping malls, hotels, commercial office buildings. & C, D, G, E, F, H, I, L \\
Beijing & 2013 & Thermal power generation, cement production, heat generation and supply, other power generation, electricity supply, data centers, integrated circuit manufacturing; wastewater treatment and reuse, water supply; urban rail transit, public buses, road freight transport, taxis, postal services; petrochemicals, other services, and miscellaneous sectors. & C, D, G, H, I, O \\
Tianjin & 2013 & Steel, chemical, petrochemical, oil and gas extraction, aviation, non-ferrous metals, pharmaceutical manufacturing, machinery manufacturing, agricultural and sideline food processing, electronics manufacturing, food and beverage, mining, rubber and plastic products. & C, D, G, B \\
Hubei & 2014 & Heat generation and supply, cement, textile industry, chemical industry, non-ferrous and other metal products, food and beverage, pharmaceuticals, papermaking, glass and other building materials, ceramic manufacturing, automobile manufacturing, equipment manufacturing, steel, petrochemicals, water supply, and other industries. & C, D, G \\
Chongqing & 2014 & Automobile manufacturing, electronics manufacturing, pharmaceuticals; other non-ferrous metal smelting and rolling; food, tobacco, alcohol, beverage and tea production; glass and glass products manufacturing; papermaking; ceramics; oil and gas; cement grinding process; machinery manufacturing; other industrial sectors; chemical industry, steel industry, flat glass, petrochemicals. & C, D, G \\
Sichuan & 2016 & Power generation, petrochemicals, building materials, steel, non-ferrous metals, and other energy-intensive industries. & C, D, G \\
Fujian & 2016 & Power generation, steel, chemical, petrochemicals, non-ferrous metals, civil aviation, building materials, papermaking, ceramics. & C, D, G \\
\bottomrule
\end{tabular}
\label{Appendix-carbon indus}
\end{table}

%% file: Tables.tex/Appendix-energy_indus.tex
\begin{table}[htbp]
\centering
\caption{Energy-use rights trading pilots and coverage.}
\scriptsize
\renewcommand{\arraystretch}{1.4}
\begin{tabular}{p{3.2cm}p{1.5cm}p{9.5cm}p{1.5cm}}
\toprule
\textbf{Pilots} & \textbf{Start year} & \textbf{Coverage (summarized from official documents)} & \textbf{Industry classification codes} \\
\midrule
Henan & 2016 & Key energy-consuming enterprises (industrial enterprises) with an annual total energy consumption of 5{,}000 tons of standard coal. & All \\
Zhejiang & 2016 & Energy use trading participants include municipal and county-level governments and relevant enterprises. & All \\
Fujian & 2016 & Energy users include those required to participate in the energy use trading system and those that participate voluntarily. & All \\
Sichuan & 2016 & Key energy-using entities are provisionally defined as enterprises and institutions within the province with an annual total energy consumption of 10{,}000 tons of standard coal equivalent or more (including equivalent forms). & All \\
\bottomrule
\end{tabular}
\label{Appendix-energy indus}
\end{table}

%% file: Tables.tex/Appendix-electricity_indus.tex
\begin{table}[htbp]
\centering
\caption{Green electricity trading pilots and coverage.}
\scriptsize
\renewcommand{\arraystretch}{1.4}
\begin{tabular}{p{3.6cm}p{1.6cm}p{9.2cm}p{2.1cm}}
\toprule
\textbf{Pilots} & \textbf{Start year} & \textbf{Coverage (summarized from official documents)} & \textbf{Industry classification codes} \\
\midrule
Beijing & 2021 & Market participants include power generation enterprises (initially focused on renewable energy companies such as wind and solar power), electricity users (those with green electricity consumption and certification needs, willing to bear the environmental value of green electricity), power retailers, and grid companies. & All \\
Guangdong & 2021 & Market participants include power generation enterprises (initially focused on renewable energy companies such as wind and solar power), electricity users (including those purchasing electricity via the power market, self-generation enterprises, entities bearing consumption responsibility weight, including both total and non-hydro responsibility weights), power retailers, and grid companies. & All \\
Inner Mongolia & 2021 & Market participants include power generation enterprises (such as active coal-fired units in the western Inner Mongolia grid, wind and solar power projects that meet market access conditions, and those allowed to participate directly in trading), electricity users (excluding residential and agricultural users; all commercial and industrial users with voltage level of 10 kV and above are generally required to participate), power retailers, and new business entities. & All except A \\
\bottomrule
\end{tabular}
\label{Appendix-electricity indus}
\end{table}

%% file: Tables.tex/Appendix-indus_classification.tex
\begin{table}[htbp]
\centering
\caption{Code Classification of Industries and Description.}
\scriptsize
\renewcommand{\arraystretch}{1.3}
\begin{tabular}{p{1cm}p{5cm}p{7.5cm}}
\toprule
\textbf{Code} & \textbf{Industry classification} & \textbf{Description} \\
\midrule
A & Agriculture, Forestry, Animal Husbandry and Fishery & Farming, forestry, animal husbandry, aquaculture, etc. \\
B & Mining Industry & Coal, petroleum, and metal ore extraction and processing. \\
C & Manufacturing & Industrial manufacturing such as electrical, mechanical, food, pharma. \\
D & Electricity, Heat, Gas and Water Supply & Power generation, gas supply, heating, and water services. \\
E & Construction & Housing construction, civil engineering, interior and exterior works. \\
F & Wholesale and Retail Trade & Commodity wholesale, retail, automobile sales, etc. \\
G & Transportation, Storage and Postal Services & Road, rail, water, air transport, logistics, and courier services. \\
H & Accommodation and Catering Services & Hotels, restaurants, food delivery, etc. \\
I & Information Transmission, Software and IT Services & Telecommunications, internet services, software development, etc. \\
J & Financial Industry & Banking, insurance, securities, trust services, etc. \\
K & Real Estate & Real estate development and property management services. \\
L & Leasing and Business Services & Leasing, consulting, human resources outsourcing, etc. \\
M & Scientific Research and Technical Services & R\&D institutions, inspection/testing, and professional services. \\
N & Water Conservancy, Environment and Public Utilities & Water services, environmental protection, and waste treatment. \\
O & Resident Services, Repairs and Other Services & Repair services for vehicles, electronics, and household products. \\
P & Education & All types of schools and education-related services. \\
Q & Health and Social Work & Hospitals, clinics, elderly care, childcare services, etc. \\
R & Culture, Sports and Entertainment & Media, publishing, film, gaming, sports, etc. \\
S & Public Administration, Social Security and Organizations & Government agencies and social security institutions. \\
\bottomrule
\end{tabular}
\label{Appendix-indus classification}
\end{table}

%% file: multipol.bbl
\begin{thebibliography}{44}
\providecommand{\natexlab}[1]{#1}

\bibitem[{Bai(2009)}]{bai2009panel}
Bai, Jushan. 2009.
\newblock Panel data models with interactive fixed effects.
\newblock \emph{Econometrica} 77~(4): 1229--1279.

\bibitem[{Bersani et~al.(2022)Bersani, Falbo, and Mastroeni}]{bersani2022ets}
Bersani, Alberto~M, Paolo Falbo, and Loretta Mastroeni. 2022.
\newblock Is the ets an effective environmental policy? undesired interaction between energy-mix, fuel-switch and electricity prices.
\newblock \emph{Energy Economics} 110: 105981.

\bibitem[{Borusyak et~al.(2024)Borusyak, Jaravel, and Spiess}]{borusyak2024revisiting}
Borusyak, Kirill, Xavier Jaravel, and Jann Spiess. 2024.
\newblock Revisiting event-study designs: robust and efficient estimation.
\newblock \emph{Review of Economic Studies} 91~(6): 3253--3285.

\bibitem[{Callaway and Sant’Anna(2021)}]{callaway2021difference}
Callaway, Brantly, and Pedro~HC Sant’Anna. 2021.
\newblock Difference-in-differences with multiple time periods.
\newblock \emph{Journal of econometrics} 225~(2): 200--230.

\bibitem[{Carvalho et~al.(2018)Carvalho, Masini, and Medeiros}]{carvalho2018arco}
Carvalho, Carlos, Ricardo Masini, and Marcelo~C Medeiros. 2018.
\newblock Arco: An artificial counterfactual approach for high-dimensional panel time-series data.
\newblock \emph{Journal of econometrics} 207~(2): 352--380.

\bibitem[{Chay and Greenstone(2005)}]{chay2005does}
Chay, Kenneth~Y, and Michael Greenstone. 2005.
\newblock Does air quality matter? evidence from the housing market.
\newblock \emph{Journal of political Economy} 113~(2): 376--424.

\bibitem[{Chen et~al.(2022{\natexlab{a}})Chen, Oliva, and Zhang}]{chen2022effect}
Chen, Shuai, Paulina Oliva, and Peng Zhang. 2022{\natexlab{a}}.
\newblock The effect of air pollution on migration: Evidence from china.
\newblock \emph{Journal of Development Economics} 156: 102833.

\bibitem[{Chen et~al.(2022{\natexlab{b}})Chen, Li, Chen, and Huang}]{chen2022environmental}
Chen, Xiaoqi, Weiping Li, Zifang Chen, and Jiashun Huang. 2022{\natexlab{b}}.
\newblock Environmental regulation and real earnings management—evidence from the so2 emissions trading system in china.
\newblock \emph{Finance Research Letters} 46: 102418.

\bibitem[{Chen et~al.(2024)Chen, Zhang, Guo, and Ji}]{chen2024emission}
Chen, Yajie, Dayong Zhang, Kun Guo, and Qiang Ji. 2024.
\newblock Emission trading schemes and cross-border mergers and acquisitions.
\newblock \emph{Journal of Environmental Economics and Management} 124: 102949.

\bibitem[{Dechezlepr{\^e}tre et~al.(2023)Dechezlepr{\^e}tre, Nachtigall, and Venmans}]{dechezlepretre2023joint}
Dechezlepr{\^e}tre, Antoine, Daniel Nachtigall, and Frank Venmans. 2023.
\newblock The joint impact of the european union emissions trading system on carbon emissions and economic performance.
\newblock \emph{Journal of Environmental Economics and Management} 118: 102758.

\bibitem[{del R{\'\i}o~Gonz{\'a}lez(2007)}]{del2007interaction}
del R{\'\i}o~Gonz{\'a}lez, Pablo. 2007.
\newblock The interaction between emissions trading and renewable electricity support schemes. an overview of the literature.
\newblock \emph{Mitigation and adaptation strategies for global change} 12: 1363--1390.

\bibitem[{Fischer and Preonas(2010)}]{fischer2010combining}
Fischer, Carolyn, and Louis Preonas. 2010.
\newblock Combining policies for renewable energy: Is the whole less than the sum of its parts? .

\bibitem[{Goldsmith-Pinkham et~al.(2024)Goldsmith-Pinkham, Hull, and Koles{\'a}r}]{goldsmith2024contamination}
Goldsmith-Pinkham, Paul, Peter Hull, and Michal Koles{\'a}r. 2024.
\newblock Contamination bias in linear regressions.
\newblock \emph{American Economic Review} 114~(12): 4015--4051.

\bibitem[{Goodman-Bacon(2021)}]{goodman2021difference}
Goodman-Bacon, Andrew. 2021.
\newblock Difference-in-differences with variation in treatment timing.
\newblock \emph{Journal of econometrics} 225~(2): 254--277.

\bibitem[{Herrnstadt et~al.(2021)Herrnstadt, Heyes, Muehlegger, and Saberian}]{herrnstadt2021air}
Herrnstadt, Evan, Anthony Heyes, Erich Muehlegger, and Soodeh Saberian. 2021.
\newblock Air pollution and criminal activity: Microgeographic evidence from chicago.
\newblock \emph{American Economic Journal: Applied Economics} 13~(4): 70--100.

\bibitem[{Hintermann and Gronwald(2019)}]{hintermann2019linking}
Hintermann, Beat, and Marc Gronwald. 2019.
\newblock Linking with uncertainty: the relationship between eu ets pollution permits and kyoto offsets.
\newblock \emph{Environmental and Resource Economics} 74: 761--784.

\bibitem[{Huang et~al.(2025)Huang, Fang, Liu, and Guo}]{huang2025has}
Huang, Ying, Kai Fang, Gengyuan Liu, and Sujian Guo. 2025.
\newblock Has the carbon emission trading scheme induced investment leakage in china? firm-level evidence from china's stock market.
\newblock \emph{Energy Economics} 141: 108091.

\bibitem[{{International Energy Agency}(2022)}]{iea2022china}
{International Energy Agency}. 2022.
\newblock China energy system carbon neutrality roadmap.
\newblock Information notice, International Energy Agency (IEA).
\newblock Revised version. Summary available on page 3.

\bibitem[{Lanoie et~al.(1998)Lanoie, Thomas, and Fearnley}]{lanoie1998firms}
Lanoie, Paul, Mark Thomas, and Joan Fearnley. 1998.
\newblock Firms responses to effluent regulations: pulp and paper in ontario, 1985-1989.
\newblock \emph{Journal of Regulatory Economics} 13~(2): 103--120.

\bibitem[{Li et~al.(2025)Li, Zhang, and Gao}]{li2025impact}
Li, Donghui, Zhanxiang Zhang, and Xin Gao. 2025.
\newblock The impact of carbon emission trading system on the implied cost of equity capital.
\newblock \emph{International Review of Economics \& Finance} : 104157.

\bibitem[{Li and Zhu(2019)}]{li2019synergy}
Li, Yuan, and Lei Zhu. 2019.
\newblock Study on the synergistic effects between energy-saving trading and carbon market and the strategic choose of energy-intensive industries.
\newblock \emph{Journal of Industrial Technological Economics} 38~(7): 136--142.

\bibitem[{Liu et~al.(2022)Liu, Ren, and Li}]{liu2022so2}
Liu, Donghua, Shenggang Ren, and Wenming Li. 2022.
\newblock So2 emissions trading and firm exports in china.
\newblock \emph{Energy Economics} 109: 105978.

\bibitem[{Luan et~al.(2025)Luan, Liu, and Mei}]{luan2025impact}
Luan, Limin, Pengfei Liu, and Yingdan Mei. 2025.
\newblock The impact of pilot carbon market on firms' performance in china.
\newblock \emph{Energy Economics} 142: 108164.

\bibitem[{Morthorst(2001)}]{morthorst2001interactions}
Morthorst, Poul~Erik. 2001.
\newblock Interactions of a tradable green certificate market with a tradable permits market.
\newblock \emph{Energy policy} 29~(5): 345--353.

\bibitem[{Porter(1996)}]{porter1996america}
Porter, Michael. 1996.
\newblock America’s green strategy.
\newblock \emph{Business and the environment: a reader} 33: 1072.

\bibitem[{Proen{\c{c}}a and Fortes(2020)}]{proencca2020synergies}
Proen{\c{c}}a, Sara, and Patr{\'\i}cia Fortes. 2020.
\newblock The synergies between eu climate and renewable energy policies--evidence from portugal using integrated modelling.
\newblock \emph{Economics of Energy \& Environmental Policy} 9~(2): 149--164.

\bibitem[{Rogge and Reichardt(2016)}]{rogge2016policy}
Rogge, Karoline~S, and Kristin Reichardt. 2016.
\newblock Policy mixes for sustainability transitions: An extended concept and framework for analysis.
\newblock \emph{Research policy} 45~(8): 1620--1635.

\bibitem[{Sun et~al.(2024)Sun, Zhang, Zhou et~al.}]{sun2024compliance}
Sun, Jingqi, Zhuola Zhang, Zhenxi Zhou, et~al. 2024.
\newblock Quota compliance strategies of thermal power enterprises under the coordination of energy-consuming right trading and carbon emission trading market.
\newblock \emph{China Environmental Science} 44~(4): 1840--1850.

\bibitem[{Sun et~al.(2023)Sun, Zhou, Wang et~al.}]{sun2023synergy}
Sun, Jingqi, Yiquan Zhou, Yuan Wang, et~al. 2023.
\newblock Research on the synergistic effect of reducting pollution and carbon emissions under the interaction of market-based environmental regulations.
\newblock \emph{Chinese Journal of Environmental Management} 15~(2): 48--57.

\bibitem[{Tang et~al.(2023)Tang, Liu, and Zhang}]{tang2023does}
Tang, Chun, Xiaoxing Liu, and Chenyao Zhang. 2023.
\newblock Does china's green power trading policy play a role?-evidence from renewable energy generation enterprises.
\newblock \emph{Journal of Environmental Management} 345: 118775.

\bibitem[{Tol(1994)}]{tol1994damage}
Tol, Richard~SJ. 1994.
\newblock The damage costs of climate change: a note on tangibles and intangibles, applied to dice.
\newblock \emph{Energy Policy} 22~(5): 436--438.

\bibitem[{Tol(2018)}]{tol2018economic}
Tol, Richard~SJ. 2018.
\newblock The economic impacts of climate change.
\newblock \emph{Review of environmental economics and policy} .

\bibitem[{van~den Bergh et~al.(2021)van~den Bergh, Castro, Drews, Exadaktylos, Foramitti, Klein, Konc et~al.}]{van2021designing}
van~den Bergh, JCJM, Juana Castro, Stefan Drews, Filippos Exadaktylos, Jo{\"e}l Foramitti, Franziska Klein, Th{\'e}o Konc, et~al. 2021.
\newblock Designing an effective climate-policy mix: accounting for instrument synergy.
\newblock \emph{Climate Policy} 21~(6): 745--764.

\bibitem[{Wang et~al.(2021)Wang, Zhang, Su, Shen, Li, and Li}]{wang2021coordination}
Wang, Ge, Qi~Zhang, Bin Su, Bo~Shen, Yan Li, and Zhengjun Li. 2021.
\newblock Coordination of tradable carbon emission permits market and renewable electricity certificates market in china.
\newblock \emph{Energy Economics} 93: 105038.

\bibitem[{Wang et~al.(2025)Wang, Liu, and Zhou}]{wang2025towards}
Wang, Meiling, Zichen Liu, and Bingxuan Zhou. 2025.
\newblock Towards carbon neutrality: The impact of energy right trading policy on carbon performance of manufacturing enterprises.
\newblock \emph{Energy} 323: 135858.

\bibitem[{Wang et~al.(2024)Wang, Wang, and Wu}]{wang2024impact}
Wang, Weilong, Jianlong Wang, and Haitao Wu. 2024.
\newblock The impact of energy-consuming rights trading on green total factor productivity in the context of digital economy: Evidence from listed firms in china.
\newblock \emph{Energy Economics} 131: 107342.

\bibitem[{Wei et~al.(2023)Wei, An, and Tu}]{wei2023interaction}
Wei, Qingpo, Gang An, and Yongqian Tu. 2023.
\newblock Interaction mechanism and empirical study between carbon trading market and green electricity policies.
\newblock \emph{China Soft Science} ~(5): 198--206.

\bibitem[{Wilts and O'Brien(2019)}]{wilts2019policy}
Wilts, Henning, and Meghan O'Brien. 2019.
\newblock A policy mix for resource efficiency in the eu: key instruments, challenges and research needs.
\newblock \emph{Ecological economics} 155: 59--69.

\bibitem[{Wu et~al.(2024)Wu, Zhou, and Zha}]{wu2024interplay}
Wu, Changsong, Dequn Zhou, and Donglan Zha. 2024.
\newblock The interplay of the carbon market, the tradable green certificate market, and electricity market in south korea: Dynamic transmission and spillover effects.
\newblock \emph{Energy \& Environment} 35~(1): 163--184.

\bibitem[{Xu(2017)}]{xu2017generalized}
Xu, Yiqing. 2017.
\newblock Generalized synthetic control method: Causal inference with interactive fixed effects models.
\newblock \emph{Political Analysis} 25~(1): 57--76.

\bibitem[{{Yale Center for Environmental Law and Policy}(2024)}]{yale_epi2024}
{Yale Center for Environmental Law and Policy}. 2024.
\newblock Environmental performance index (epi) 2024.
\newblock \emph{Yale EPI Official Website} Available at: \url{https://epi.yale.edu/measure/2024/EPI} [Accessed 19 Jun. 2025].

\bibitem[{Zhang et~al.(2023)Zhang, Guo, and Zhang}]{zhang2023assessing}
Zhang, Xinyue, Xiaopeng Guo, and Xingping Zhang. 2023.
\newblock Assessing the policy synergy among power, carbon emissions trading and tradable green certificate market mechanisms on strategic gencos in china.
\newblock \emph{Energy} 278: 127833.

\bibitem[{Zhou et~al.(2022)Zhou, Xin, and Li}]{zhou2022assessing}
Zhou, Anhua, Ling Xin, and Jun Li. 2022.
\newblock Assessing the impact of the carbon market on the improvement of china's energy and carbon emission performance.
\newblock \emph{Energy} 258: 124789.

\bibitem[{Zhu and Yu(2023)}]{zhu2023coeffect}
Zhu, Siyu, and Bing Yu. 2023.
\newblock Research on the co-benefits of pollution reduction and carbon reduction of “emissions trading” and “carbon emissions trading”——based on the dual perspectives of pollution control and policy management.
\newblock \emph{Chinese Journal of Environmental Management} 15~(1): 102--109.

\end{thebibliography}
